\documentclass[12pt,a4paper]{article}
\usepackage{graphicx}
\usepackage{subeqnarray}
\begin{document}
%
%
%
%
\def\astrobj#1{#1}
\newenvironment{lefteqnarray}{\arraycolsep=0pt\begin{eqnarray}}
{\end{eqnarray}\protect\aftergroup\ignorespaces}
\newenvironment{lefteqnarray*}{\arraycolsep=0pt\begin{eqnarray*}}
{\end{eqnarray*}\protect\aftergroup\ignorespaces}
\newenvironment{leftsubeqnarray}{\arraycolsep=0pt\begin{subeqnarray}}
{\end{subeqnarray}\protect\aftergroup\ignorespaces}
\newcommand{\diff}{{\rm\,d}}
\newcommand{\pprime}{{\prime\prime}}
\newcommand{\szeta}{\mskip 3mu /\mskip-10mu \zeta}
\newcommand{\FC}{\mskip 0mu {\rm F}\mskip-10mu{\rm C}}
\newcommand{\appleq}{\stackrel{<}{\sim}}
\newcommand{\appgeq}{\stackrel{>}{\sim}}
\newcommand{\legr}{\stackrel{<}{>}}
\newcommand{\grle}{\stackrel{>}{<}}
\newcommand{\Int}{\mathop{\rm Int}\nolimits}
\newcommand{\Nint}{\mathop{\rm Nint}\nolimits}
\newcommand{\range}{{\rm -}}
\newcommand{\displayfrac}[2]{\frac{\displaystyle #1}{\displaystyle #2}}
\def\astrobj#1{#1}
%
\title{
Fractional yields inferred from halo and thick disk stars}
\author{{$\qquad~$R.~Caimmi}\footnote{
{\it Physics and Astronomy Department, Padua Univ., Vicolo Osservatorio 3/2,
I-35122 Padova, Italy}
email: roberto.caimmi@unipd.it~~~
fax: 39-049-8278212}
\phantom{agga}}
%
%
\maketitle
\begin{quotation}
\section*{}
\begin{Large}
\begin{center}
Abstract

\end{center}
\end{Large}
\begin{small}

\noindent\noindent

Linear [Q/H]-[O/H] relations, Q = Na, Mg, Si, Ca, Ti, Cr, Fe, Ni, are inferred
from a sample $(N=67)$ of recently studied FGK-type dwarf stars in the solar
neighbourhood including different populations (Nissen and Schuster 2010;
Ramirez et al. 2012), namely LH
($N=24$, low-$\alpha$ halo), HH ($N=25$, high-$\alpha$ halo), KD ($N=16$,
thick disk), OL ($N=2$, globular cluster outliers).   Regression line slope
and intercept estimators and related variance estimators
are determined.   With regard to the
straight line, [Q/H]=$a_{\rm Q}$[O/H]+$b_{\rm Q}$, sample stars display along
a ``main sequence'', [Q,O] = [$a_{\rm Q},b_{\rm Q},\Delta b_{\rm Q}$], leaving
aside the two OL stars which, in most cases (e.g.,
Na), lie outside.   A unit slope, $a_{\rm Q}=1$, implies Q is a primary
element synthesised via SNII progenitors in presence of universal
stellar initial mass function (defined as simple primary element).
To this respect, Mg, Si, Ti, show $\hat a_{\rm Q}=1$ within $\mp2\hat\sigma_
{\hat a_{\rm Q}}$; Cr, Fe, Ni, within $\mp3\hat\sigma_{\hat a_{\rm Q}}$; Na,
Ca, within $\mp r\hat\sigma_{\hat a_{\rm Q}}$, $r>3$.
The empirical, differential element abundance distributions are inferred from
LH, HH, KD, HA = HH + KD subsamples, where related regression lines represent
their theoretical counterparts within the framework of simple MCBR (multistage
closed box + reservoir) chemical evolution models.  Hence the fractional
yields, $\hat{p}_{\rm Q}/\hat{p}_{\rm O}$, are determined and (as an example)
a comparison is shown with their theoretical counterparts inferred from SNII
progenitor nucleosynthesis under the assumption of a power-law stellar initial
mass function.   The generalized
fractional yields, $C_{\rm Q}=Z_{\rm Q}/Z_{\rm O}^{a_{\rm Q}}$, are determined
regardless of the chemical evolution model.
%
%
The ratio of outflow to star formation rate is compared for different
populations, in the framework of simple MCBR models.   The opposite situation
of element abundance variation entirely due to cosmic scatter is also
considered under
reasonable assumptions.   The related differential element abundance
distribution fits to the data as well as its counterpart inferred in the
opposite limit of instantaneous mixing in presence of chemical evolution,
while the latter is preferred for HA subsample.

\noindent
{\it keywords -
Galaxy: evolution - Galaxy: formation - stars: evolution - stars: formation.}
\end{small}
\end{quotation}

\section{Introduction} \label{intro}

Leaving aside the first three minutes after the birth of the universe,
elements heavier than He, or metals, are synthesised within stars and
returned to the interstellar medium via supernova (SN) explosions, with the
addition of envelope loss from planetary nebulae.   There are two main types
of SN. SNII progenitors are main-sequence stars, massive $(m\appgeq10m_\odot)$
and short-lived $(0.001\appleq\tau/{\rm Gyr}\appleq0.01)$, producing a wide
variety of nuclides among which are $\alpha$ elements (traditionally, Mg, Si,
Ca, Ti, withthe addition of O) and Fe (e.g., Woosley and Weaver 1995;
Kobayashi et al. 2011). SNIa progenitors are white dwarfs belonging to a
binary system, less massive $(m\appleq1.4m_\odot)$, generally long-living
$(\tau\appgeq1 {\rm Gyr})$, which mainly produce Fe (e.g., Kobayashi et al.
1998; Kobayashi and Nomoto 2009).

Element production within stars has been a major research focus for many years
concerning e.g., abundance ratios as a function of the metallicity (Wheeler et
al. 1989), nucleosynthesis in massive $(11\le m/m_\odot\le40)$ stars with
solar and subsolar metal abundance (Woosley and Weaver 1995), chemical
evolution of the solar neighbourhhod for elements up to zinc (Timmes et al.
1995), $\alpha$ element production and comparison with the data from different
populations (MacWilliam 1997), stellar evolution including close binaries
(Wallerstein et al. 1997), chemical evolution of Galactic and extra Galactic
populations (Venn et al. 2004), evolution of the isotope ratios of elemental
abundances (from C to Zn) in different Galactic populations (Kobayashi et al.
2011).

The fractional logarithmic number abundance or, in short, number abundance,
[Q/Fe], where Q denotes a generic nuclide and, in particular, an
$\alpha$ element, has been the subject of several investigations (e.g.,
Edvardsson et al. 1993; Nissen and
Schuster 1997; Fulbright 2002; Stephens and Boesgaard 2002; Gratton et al.
2003) for establishing if the distribution of [Q/Fe] in different populations
is continuous or bimodal.   Less attention, however, has been devoted to the
connection between number abundances, [Q/H], which are related to the chemical
evolution of a single element, Q, instead of a pair, Q$_1$, Q$_2$, for fixed
hydrogen abundance (e.g., Caimmi 2013; Carretta 2013).

Dealing with $\{{\sf O}[{\rm Q_1/H}][{\rm Q_2/H}]\}$ plane instead of
$\{{\sf O}[{\rm Q_1/H}][{\rm Q_2/Q_1}]\}$ could be due to two orders of
reasons, namely (i) avoiding that uncertainties in Q$_1$ reflect on both axes
and (ii) fully exploiting the different sites of nucleosynthesis for Q$_1$.
For instance, a simple [Na/H]-[Fe/H] linear relation is expected keeping in
mind Na is mainly produced via hydrostatic C-burning within SNII progenitors,
proportionally to the initial metallicity of the parent star (e.g., Woosley
and Weaver 1995).   On the other hand, additional production could take place
via proton-capture on Ne in H-burning at high temperature, which could explain
Na overabundance detected in some globular cluster stars.   For further
details and additional references, an interested reader is addressed to a
recent attempt (Carretta 2013).

According to the standard definition, a nuclide is primary when the yield is
independent of the initial composition of the parent star and secondary if
otherwise (e.g., Pagel and Tautvaisiene 1995).
Let simple primary elements be defined as synthesised via SNII
progenitors in presence of universal stellar initial mass
function.   Accordingly, the yield ratio of two
selected simple primary elements, or fractional yield, maintains constant in
time, which implies $Z_{\rm Q}/Z=(Z_{\rm Q})_\odot/Z_\odot$, $Z=\sum
Z_{\rm Q}$.

Within the framework of MCBR (multistage closed box + reservoir)
models (Caimmi 2011a; 2012a) in the linear limit (hereafter quoted as simple
MCBR models), which well hold for simple primary elements, the
fractional yield may be expressed by a short formula and then inferred from
the data related to early populations, such as the halo and the
low-metallicity ([Fe/H] $<-0.6$) thick disk.
Star formation therein spanned for less than about 1 Gyr, implying an
interstellar medium mainly enriched by SNII progenitors. Number abundances
of several elements, namely O, Na, Mg, Si, Ca, Ti, Cr, Fe, Ni, can be inferred
from recently studied samples of solar neighbourhood FGK-type dwarf stars
(Nissen and Schuster 2010, hereafter quoted as NS10; Ramirez et al. 2012,
hereafter quoted as Ra12).

The current note is devoted to (i) analysis of
the dependence of [Q/H] on [O/H], where Q $\ne$ O is any among the elements
mentioned above.   To this aim, a general classification is introduced and
constraints on the chemical evolution of related populations are inferred;
(ii) determination of empirical, differential element abundance distribution
from different subsamples, together with related theoretical counterparts in
the framework of simple MCBR models;
(iii) evaluation of fractional yields and related parameters in the framework
of simple MCBR models, including an example of comparison with theoretical
counterparts, inferred from SNII progenitor nucleosynthesis under the
assumption of power-law stellar initial mass function; (iv) determination of
theoretical, differential element abundance distribution in the opposite limit
of inhomogeneous mixing due to cosmic scatter obeying a Gaussian distribution
where the mean and the variance are evaluated from related subsamples.

Basic informations on the data (NS10; Ra12) are provided in Section 2.   The
inferred [Q/H]-[O/H] relations are shown and classified in Section 3.   The
results are discussed in Section 4.   The conclusion is presented in Section
5. Further details are illustrated in the Appendix.

\section{The data} \label{data}

The data are taken from a sample $(N=67)$ of solar neighbourhood FGK-type
dwarf
stars in the metallicity range, $-1.6<$ [Fe/H] $<-0.4$, for which [O/H] has
been determined via a non-LTE analysis of the 777 nm OI triplet lines (Ra12),
while [Fe/H] and [Q/Fe], Q = Na, Mg, Si, Ca, Ti, Cr, Ni, are already known
from an earlier attempt (NS10).   Subsamples are extracted from the parent
sample according to different populations,
as LH (low-[$\alpha$/Fe] halo stars), HH (high-[$\alpha$/Fe] halo stars), KD
(thick disk stars), OL (globular cluster outliers).   The related population
is LH $(N=24)$, HH $(N=25)$, KD $(N=16)$, OL $(N=2)$, respectively.   KD stars
exhibit low abundances, [Fe/H]$~<-0.6$.   OL stars are included for
completeness and to get a first idea on the trend shown with respect to the
empirical [Q/H]-[O/H] relation inferred for LH, HH, KD stars.

The fractional number abundances, [O/H] and [Fe/H], are taken from related
parent papers (NS10; Ra12) while the remaining are inferred from the parent
paper (NS10) according to the standard relation, [Q/H] = [Q/Fe] + [Fe/H],
which completes the set of needed data.    For further details and exhaustive
presentation, an interested reader is referred to the parent papers (NS10;
Ra12).

For sake of simplicity, low/high-[$\alpha$/Fe] halo stars shall be quoted in
the following as low/high-$\alpha$ halo stars, where ``low/high-$\alpha$'' has
to be intended with respect to fixed [Fe/H] (e.g., NS10; Conroy 2012).

\section{Results} \label{resu}

\begin{figure}[t]  
\begin{center}      
\includegraphics[scale=0.8]{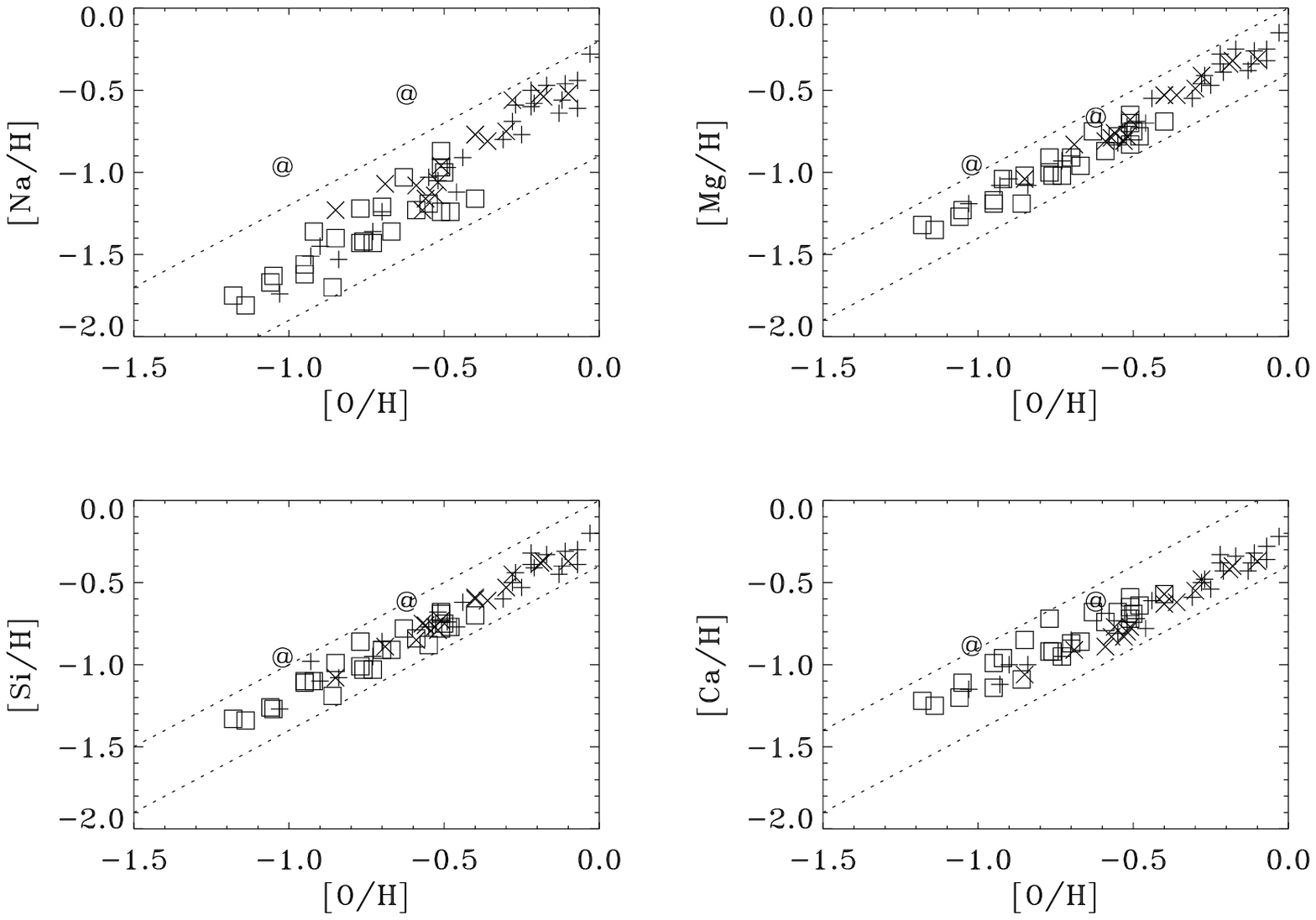}                      
\caption[ddbb]{The empirical [Q/H]-[O/H] relation, Q = Na, Mg, Si, Ca, for
subsamples, LH (low-$\alpha$ halo stars, open squares), HH (high-$\alpha$ halo
stars, crosses), KD (low-metallicity thick disk stars, saltires), OL (globular
cluster
outliers, ``at'' symbols).   Also shown for comparison is the narrowest ``main
sequence'', [Q,O] = $[1,b_{\rm Q},\Delta b_{\rm Q}]$,
within which the data lie leaving outside OL stars.   Typical error
bars are about twice the symbol dimensions.   See text for further
details.}
\label{f:NaMgSiCa}     
\end{center}       
\end{figure}

\begin{figure}[t]  
\begin{center}      
\includegraphics[scale=0.8]{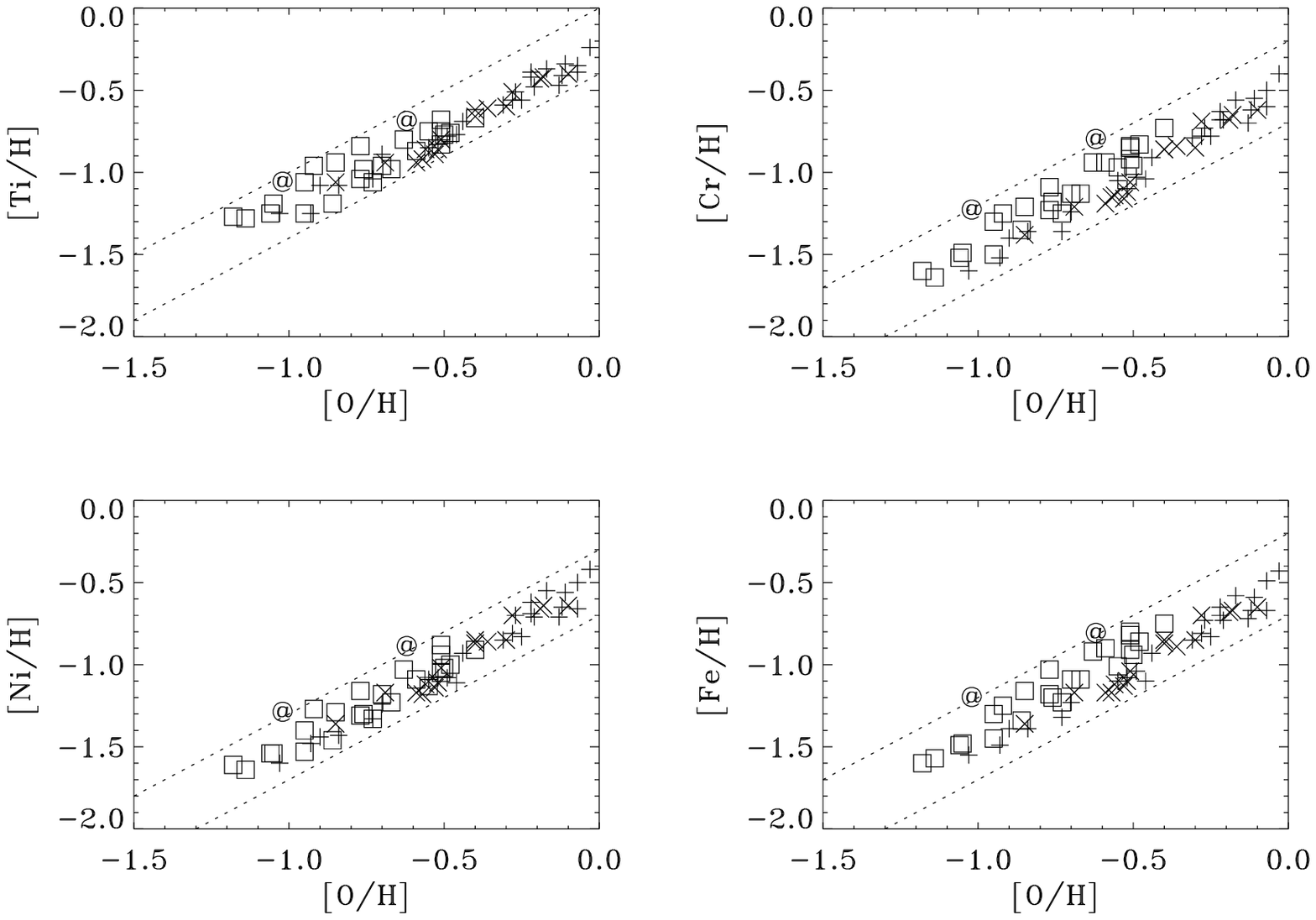}                      
\caption[ddbb]{As in Fig.\,\ref{f:NaMgSiCa}, but for Q = Ti, Cr, Fe, Ni.
}
\label{f:TiCrNiFe}     
\end{center}       
\end{figure}

Oxygen is the most abundant metal in the universe and is mainly synthesised
within SNII progenitors.   For this reason, oxygen abundance is chosen here
as reference abundance.   The empirical [Q/H]-[O/H] relations, Q = Na, Mg, Si,
Ca; Q = Ti, Cr, Fe, Ni; are plotted in Figs.\,\ref{f:NaMgSiCa} and
\ref{f:TiCrNiFe}, respectively, for LH, HH, KD, OL, subsamples.
Also shown are the narrowest ``main sequences'',
[Q,O] = $[1,b_{\rm Q},\Delta b_{\rm Q}]$, limited by the
straight lines of slope, $a_{\rm Q}=1$, and intercepts,
$b_{\rm Q}\mp\Delta b_{\rm Q}/2$, within which the data
lie leaving outside OL stars.   For two selected elements, Q$_1$ and
Q$_2$, a general classification reads
${\rm [Q_1,Q_2]}=[a_{\rm Q_1},b_{\rm Q_1},\Delta b_{\rm Q_1}]$ (Caimmi 2013).
Though OL stars are far from the main sequence only for Na, still a
similar trend is shown for the remaining elements too.

The dispersion of the data around a straight line of fixed slope can be
evaluated from the width of the main sequence measured on the vertical axis,
as the difference between the intercepts of related bounding straight lines,
$\Delta b_{\rm Q}$.   An
inspection of Figs.\,\ref{f:NaMgSiCa}-\ref{f:TiCrNiFe} shows the largest
dispersion is exhibited by Na, $\Delta b_{\rm Na}=0.7$ dex, followed by Cr and
Fe, $\Delta b_{\rm Cr}=\Delta b_{\rm Fe}=0.5$ dex, and the remaining elements,
$\Delta b_{\rm Mg}=\Delta b_{\rm Si}=\Delta b_{\rm Ca}=\Delta b_{\rm Ti}=
\Delta b_{\rm Ni}=0.4$ dex.

For each plot, the regression line has been determined for LH, HH, KD
subsamples using the bisector method (e.g., Isobe et al. 1990; Caimmi 2011b,
2012b) and the results are listed in Table \ref{t:rette}.
\begin{table*}
\caption{Regression line slope estimator, $\hat a_{\rm Q}$, square root of
variance estimator, $\hat\sigma_{\hat a_{\rm Q}}$, regression line intercept
estimator, $\hat b_{\rm Q}$, square root of variance estimator,
$\hat\sigma_{\hat b_{\rm Q}}$, 
generalized fractional yield, $C_{\rm Q}$, expressed by
Eq.\,(\ref{eq:CQ}), for Q = Na, Mg,
Si, Ca, Ti, Cr, Fe, Ni, with regard to different subsamples, LH (low-$\alpha$
halo stars), HH (high-$\alpha$ halo stars), KD (low-metallicity thick
disk stars), HA (high-$\alpha$ halo + low-metallicity thick disk stars).   See
text for further details.}
\label{t:rette}
\begin{center}
\begin{tabular}{lllllll} \hline
\multicolumn{1}{c}{Q} &
\multicolumn{1}{c}{$\hat a_{\rm Q}$} &
\multicolumn{1}{c}{$\hat\sigma_{\hat a_{\rm Q}}$} &
\multicolumn{1}{c}{$-\hat b_{\rm Q}$} &
\multicolumn{1}{c}{$\hat\sigma_{\hat b_{\rm Q}}$} &
\multicolumn{1}{c}{$C_{\rm Q}$} &
\multicolumn{1}{c}{pop} \\
\hline
Na & 1.1467D$+$00 & 1.1017D$-$01 & 4.9224D$-$01 & 9.6151D$-$02 & 3.4952D$-$03 & LH \\
   & 1.3461D$+$00 & 5.1060D$-$02 & 3.3485D$-$01 & 2.9875D$-$02 & 1.4034D$-$02 & HH \\
   & 1.3036D$+$00 & 1.4981D$-$01 & 3.1286D$-$01 & 5.4564D$-$02 & 1.1857D$-$02 & KD \\
   & 1.3339D$+$00 & 5.4134D$-$02 & 3.2406D$-$01 & 2.5128D$-$02 & 1.3508D$-$02 & HA \\
Mg & 9.3798D$-$01 & 4.8701D$-$02 & 2.6995D$-$01 & 4.1267D$-$02 & 4.8194D$-$02 & LH \\
   & 1.0333D$+$00 & 3.4714D$-$02 & 1.6663D$-$01 & 1.9884D$-$02 & 9.9892D$-$02 & HH \\
   & 1.0916D$+$00 & 5.2829D$-$02 & 1.4288D$-$01 & 2.3186D$-$02 & 1.4250D$-$01 & KD \\
   & 1.0464D$+$00 & 3.0344D$-$02 & 1.6189D$-$01 & 1.5833D$-$02 & 1.0805D$-$01 & HA \\
Si & 9.1887D$-$01 & 4.1941D$-$02 & 2.8057D$-$01 & 3.5713D$-$02 & 4.0039D$-$02 & LH \\
   & 9.7167D$-$01 & 4.4229D$-$02 & 2.2499D$-$01 & 2.1505D$-$02 & 5.9734D$-$02 & HH \\
   & 1.0258D$+$00 & 4.1416D$-$02 & 2.0678D$-$01 & 2.1943D$-$02 & 8.2335D$-$02 & KD \\
   & 9.8441D$-$01 & 3.5638D$-$02 & 2.2186D$-$01 & 1.7139D$-$02 & 6.4237D$-$02 & HA \\
Ca & 9.3394D$-$01 & 4.5183D$-$02 & 1.7341D$-$01 & 3.5430D$-$02 & 5.3438D$-$03 & LH \\
   & 9.3210D$-$01 & 3.2634D$-$02 & 2.3770D$-$01 & 1.7615D$-$02 & 4.5650D$-$03 & HH \\
   & 1.0244D$+$00 & 4.4200D$-$02 & 2.3114D$-$01 & 1.9383D$-$02 & 7.4573D$-$03 & KD \\
   & 9.5927D$-$01 & 3.2646D$-$02 & 2.3965D$-$01 & 1.4921D$-$02 & 5.2271D$-$03 & HA \\
Ti & 8.7152D$-$01 & 5.2099D$-$02 & 3.0866D$-$01 & 3.7487D$-$02 & 1.3807D$-$04 & LH \\
   & 1.0007D$+$00 & 3.5937D$-$02 & 2.6133D$-$01 & 1.7091D$-$02 & 2.9955D$-$04 & HH \\
   & 1.0667D$+$00 & 7.5124D$-$02 & 2.5133D$-$01 & 2.8450D$-$02 & 4.3066D$-$04 & KD \\
   & 1.0184D$+$00 & 3.3349D$-$02 & 2.6137D$-$01 & 1.4353D$-$02 & 3.2808D$-$04 & HA \\
Cr & 1.1622D$+$00 & 4.5627D$-$02 & 2.8681D$-$01 & 3.3824D$-$02 & 3.4549D$-$03 & LH \\
   & 1.1459D$+$00 & 3.5077D$-$02 & 4.4373D$-$01 & 1.8743D$-$02 & 2.2135D$-$03 & HH \\
   & 1.1823D$+$00 & 7.0671D$-$02 & 4.4678D$-$01 & 3.2115D$-$02 & 2.6505D$-$03 & KD \\
   & 1.1569D$+$00 & 3.2707D$-$02 & 4.4658D$-$01 & 1.6126D$-$02 & 2.3264D$-$03 & HA \\
Fe & 1.1311D$+$00 & 4.2032D$-$02 & 2.9019D$-$01 & 3.6276D$-$02 & 2.2724D$-$01 & LH \\
   & 1.0849D$+$00 & 3.9103D$-$02 & 4.8509D$-$01 & 2.3072D$-$02 & 1.1430D$-$01 & HH \\
   & 1.1035D$+$00 & 6.7964D$-$02 & 4.7888D$-$01 & 3.1622D$-$02 & 1.2766D$-$01 & KD \\
   & 1.0892D$+$00 & 3.4539D$-$02 & 4.8406D$-$01 & 1.8746D$-$02 & 1.1715D$-$01 & HA \\
Ni & 1.0165D$+$00 & 5.3955D$-$02 & 4.7925D$-$01 & 4.3627D$-$02 & 4.4933D$-$03 & LH \\
   & 1.1406D$+$00 & 3.7857D$-$02 & 4.6429D$-$01 & 2.3555D$-$02 & 8.8122D$-$03 & HH \\
   & 1.1400D$+$00 & 7.4321D$-$02 & 4.5788D$-$01 & 3.2996D$-$02 & 8.9152D$-$03 & KD \\
   & 1.1396D$+$00 & 3.4006D$-$02 & 4.6208D$-$01 & 1.8954D$-$02 & 8.8146D$-$03 & HA \\
\hline
\end{tabular}                     
\end{center}                      
\end{table*}                       
The same has been done for the HA = HH + KD subsample $(N=41)$ for exploiting
the possibility of an inner halo-thick disk chemical evolution.   An
inspection of Table \ref{t:rette} shows the following.
\begin{description}
\item[(1)\hspace{2.0mm}]
Regression line slope estimators, $\hat a_{\rm Q}$, for different populations,
are consistent within about $\mp2\hat\sigma_{\hat a_{\rm Q}}$, with the
exception of Fe where they agree within $\mp\hat\sigma_{\hat a_{\rm Q}}$.
\item[(2)\hspace{2.0mm}]
For a fixed element, regression line slope estimators may be consistent with
the unit slope within $\mp\hat\sigma_{\hat a_{\rm Q}}$ for all populations
(Si) or some (Ti) or only one (Na, Mg, Ca, Fe, Ni) or none (Cr).   For all
elements, regression line slope estimators  may be consistent with the unit
slope, regardless of the population, within $\mp2\hat\sigma_{\hat a_{\rm Q}}$
(Mg, Si, Ti) or $\mp3\hat\sigma_{\hat a_{\rm Q}}$ (Cr, Fe, Ni) or not at all
i.e. $\mp r\hat\sigma_{\hat a_{\rm Q}}$, $r>3$ (Na, Ca).
\item[(3)\hspace{2.0mm}]
Regression line intercept estimators, $\hat b_{\rm Q}$, for different
populations, may be consistent within $\mp\hat\sigma_{\hat b_{\rm Q}}$ (Ti,
Ni) or $\mp2\hat\sigma_{\hat b_{\rm Q}}$ (Na, Mg, Si, Ca) or marginally
consistent within about $\mp3\hat\sigma_{\hat b_{\rm Q}}$ (Cr, Fe).
\end{description}
In conclusion, number abundances plotted in
Figs.\,\ref{f:NaMgSiCa}-\ref{f:TiCrNiFe} show a linear trend as:
\begin{lefteqnarray}
\label{eq:QHOH}
&& {\rm[Q/H]}=a_{\rm Q}{\rm[O/H]}+b_{\rm Q}~~;
\end{lefteqnarray}
for LH, HH, KD, HA populations, according to Table \ref{t:rette}.   While
different populations may be connected with the same regression line to a
first extent, the consistency with the unit slope appears problematic in
several cases.

By definition, [O/Q] = [O/H] $-$ [Q/H], which via
Eq.\,(\ref{eq:QHOH}) translates into:
\begin{lefteqnarray}
\label{eq:OQOH}
&& {\rm[O/Q]}=(1-a_{\rm Q}){\rm[O/H]}-b_{\rm Q}~~;
\end{lefteqnarray}
where, in particular, low-[O/Q] stars relate to larger $a_{\rm Q}$ and/or
$b_{\rm Q}$ with respect to high-[O/Q] stars and, in addition, a constant
[O/Q] abundance ratio relates to the unit slope, $a_{\rm Q}=1$.   According to
the standard notation, [Q$_1/$Q$_2$]
$=\log(N_{\rm Q_1}/N_{\rm Q_2})-\log(N_{\rm Q_1}/N_{\rm Q_2})_\odot$, where
$N_{\rm Q}$ is the number density of the element, Q.

\begin{figure}[t]  
\begin{center}      
\includegraphics[scale=0.8]{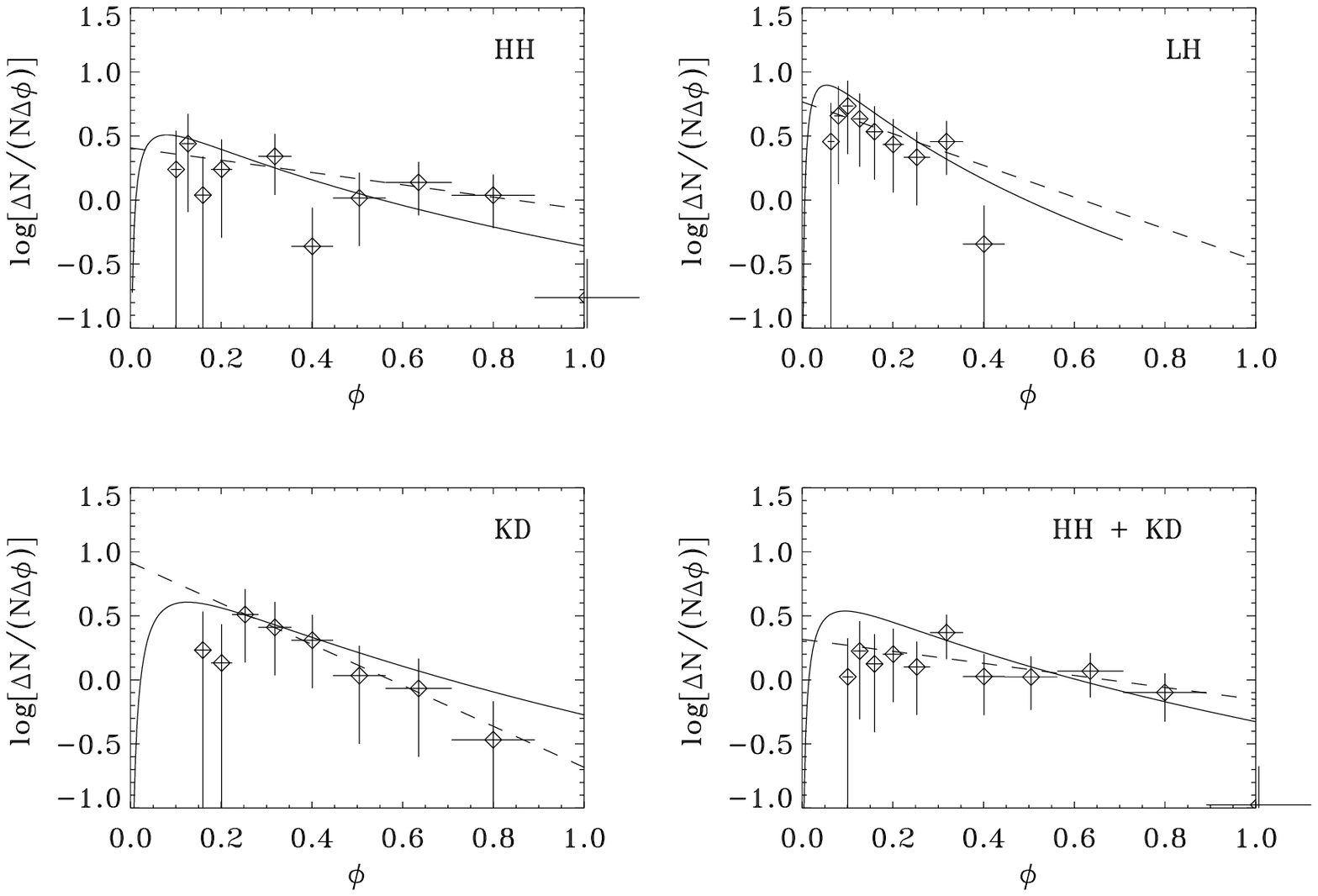}                      
\caption[ddbb]{The empirical differential oxygen abundance distribution
inferred from HH, LH, KD, HA = HH + KD subsamples.   Lower uncertainties
attaining the horizontal axis (decreasing up to negative infinity) relate to
bins populated by a single star.   Dashed straight lines represent regression
lines to points defining bins populated by at least two stars.   Full curves
represent the theoretical differential oxygen distribution due to intrinsic
scatter obeying a Gaussian distribution with mean and variance inferred from
the data.   See text for further details.}
\label{f:Ohlk4d}     
\end{center}       
\end{figure}

\begin{figure}[t]  
\begin{center}      
\includegraphics[scale=0.8]{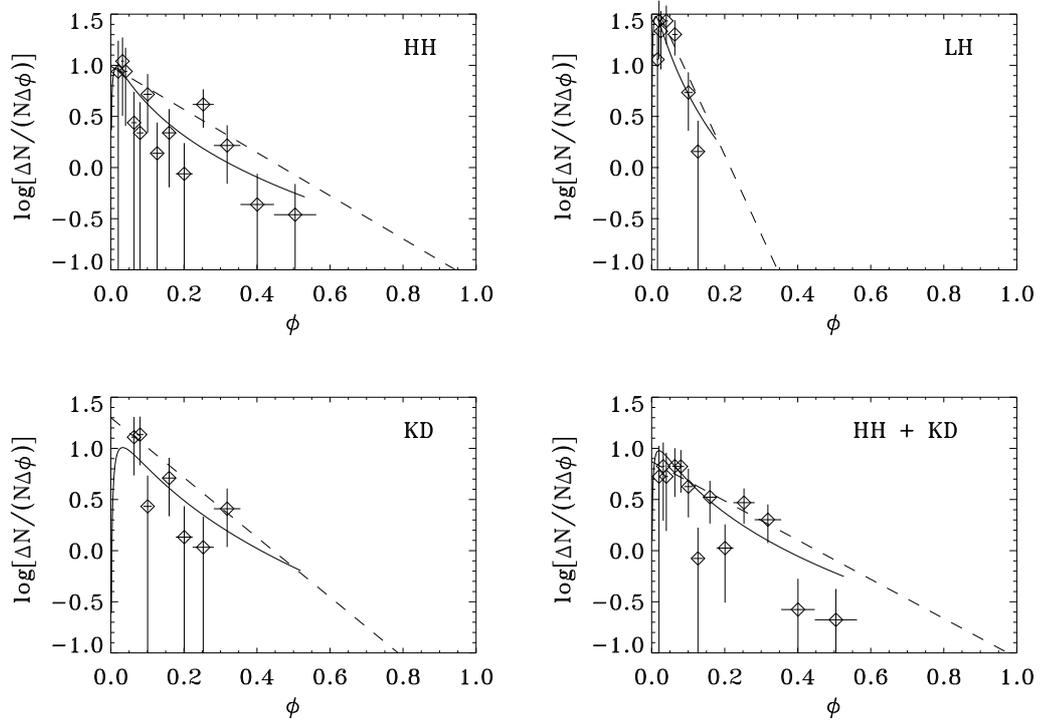}                      
\caption[ddbb]{As in Fig.\,\ref{f:Ohlk4d}, but concerning sodium instead of
oxygen.}
\label{f:Nalhk4d}     
\end{center}       
\end{figure}

\begin{figure}[t]  
\begin{center}      
\includegraphics[scale=0.8]{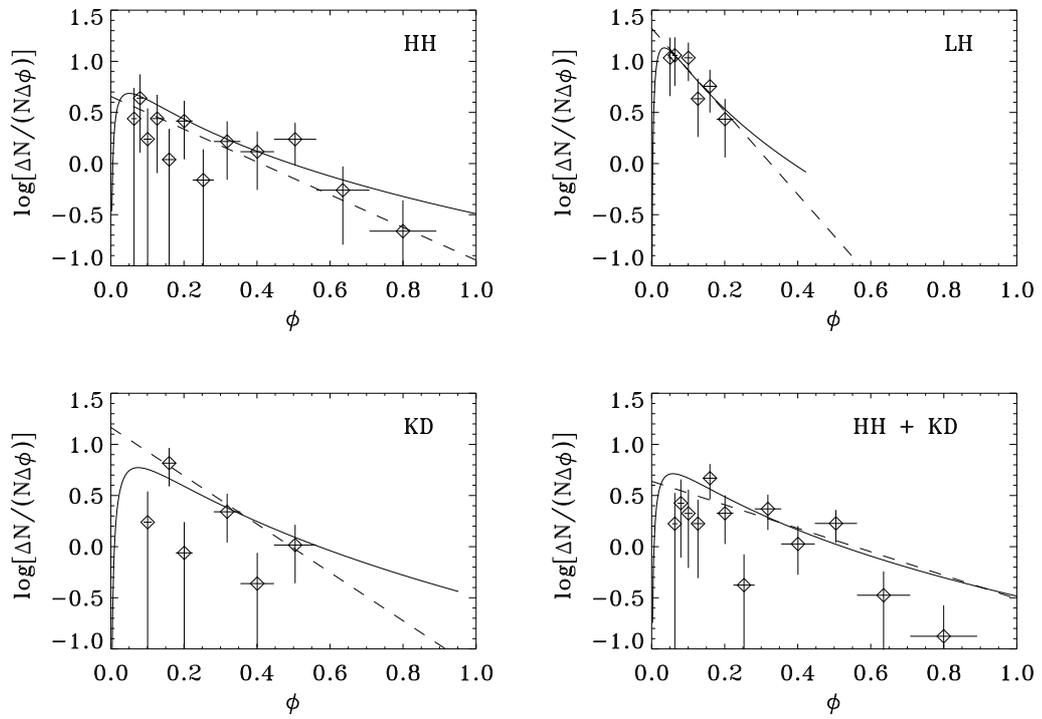}                      
\caption[ddbb]{As in Fig.\,\ref{f:Ohlk4d}, but concerning magnesium instead of
oxygen.}
\label{f:Mglhk4d}     
\end{center}       
\end{figure}

\begin{figure}[t]  
\begin{center}      
\includegraphics[scale=0.8]{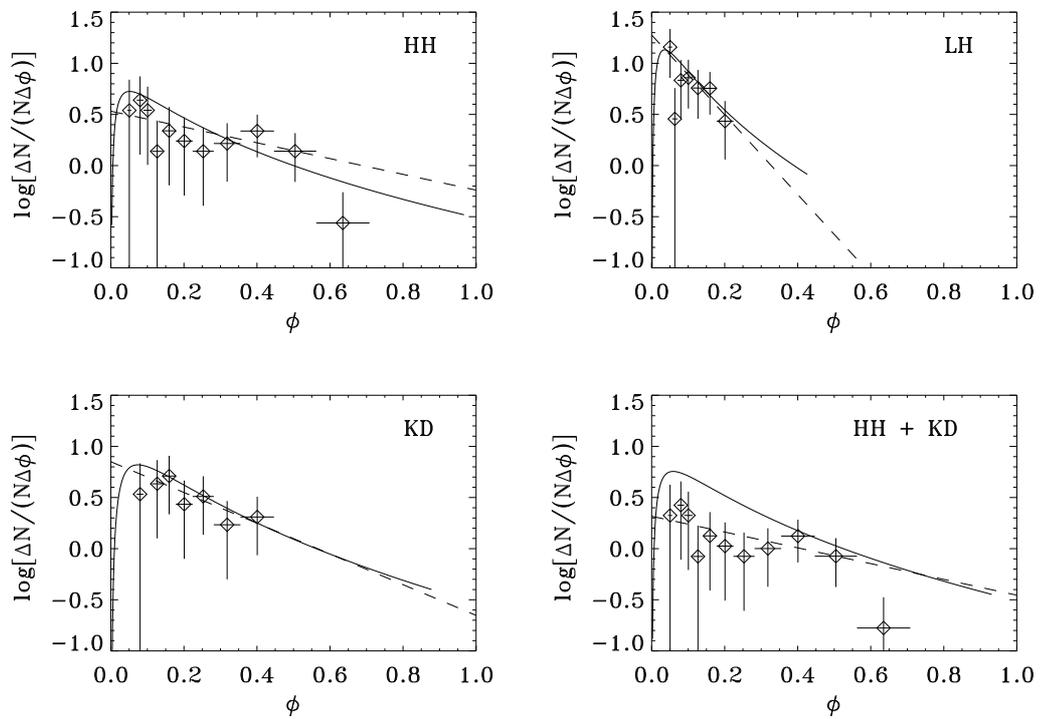}                      
\caption[ddbb]{As in Fig.\,\ref{f:Ohlk4d}, but concerning silicon instead of
oxygen.}
\label{f:Silhk4d}     
\end{center}       
\end{figure}

\begin{figure}[t]  
\begin{center}      
\includegraphics[scale=0.8]{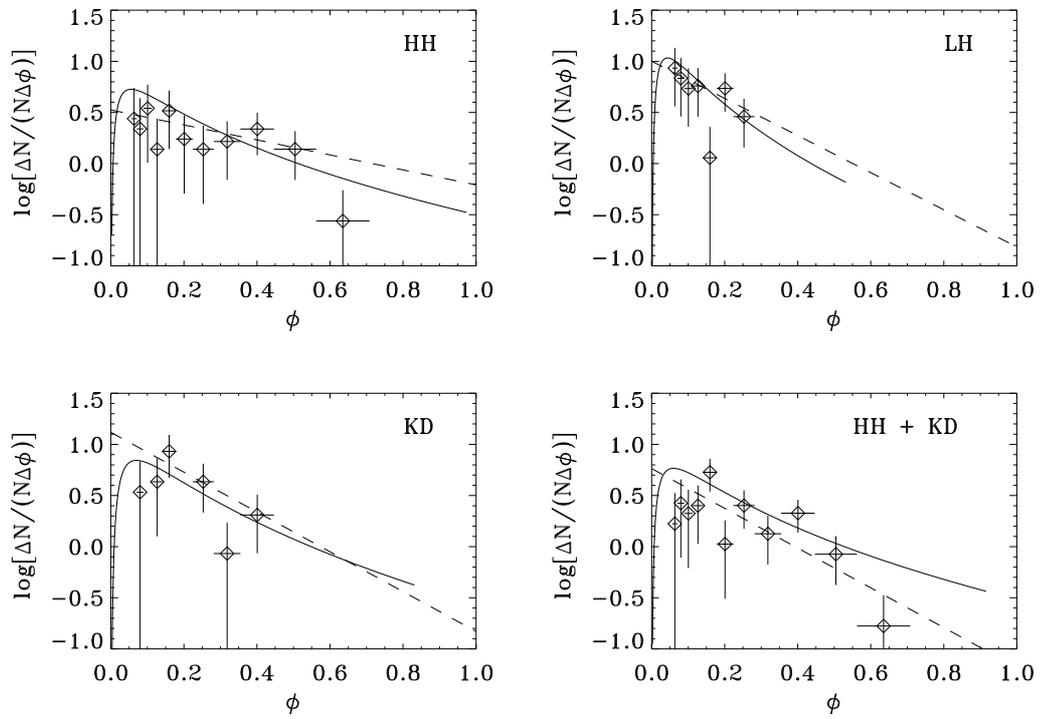}                      
\caption[ddbb]{As in Fig.\,\ref{f:Ohlk4d}, but concerning calcium instead of
oxygen.}
\label{f:Calhk4d}     
\end{center}       
\end{figure}

\begin{figure}[t]  
\begin{center}      
\includegraphics[scale=0.8]{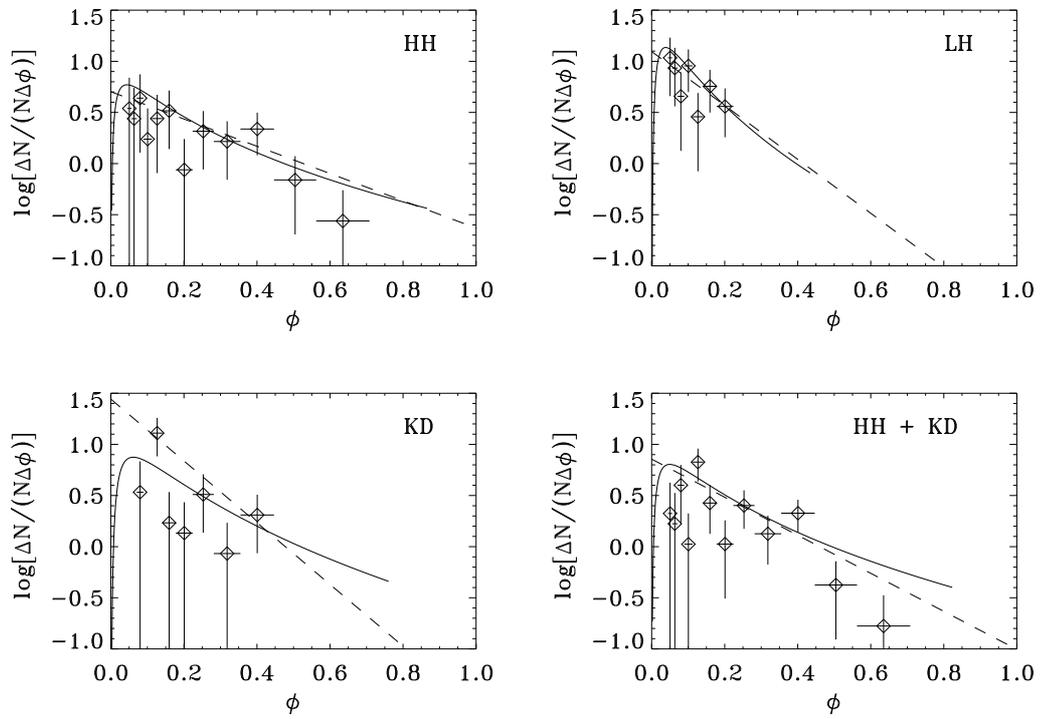}                      
\caption[ddbb]{As in Fig.\,\ref{f:Ohlk4d}, but concerning titanium instead of
oxygen.}
\label{f:Tilhk4d}     
\end{center}       
\end{figure}

\begin{figure}[t]  
\begin{center}      
\includegraphics[scale=0.8]{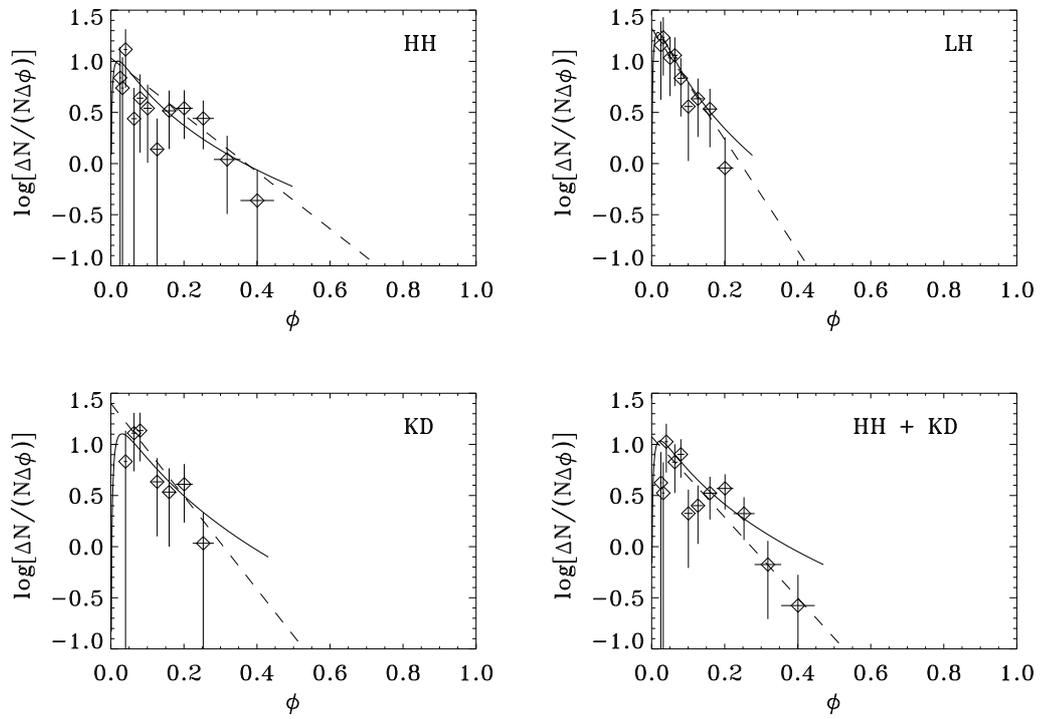}                      
\caption[ddbb]{As in Fig.\,\ref{f:Ohlk4d}, but concerning chromium instead of
oxygen.}
\label{f:Crlhk4d}     
\end{center}       
\end{figure}

\begin{figure}[t]  
\begin{center}      
\includegraphics[scale=0.8]{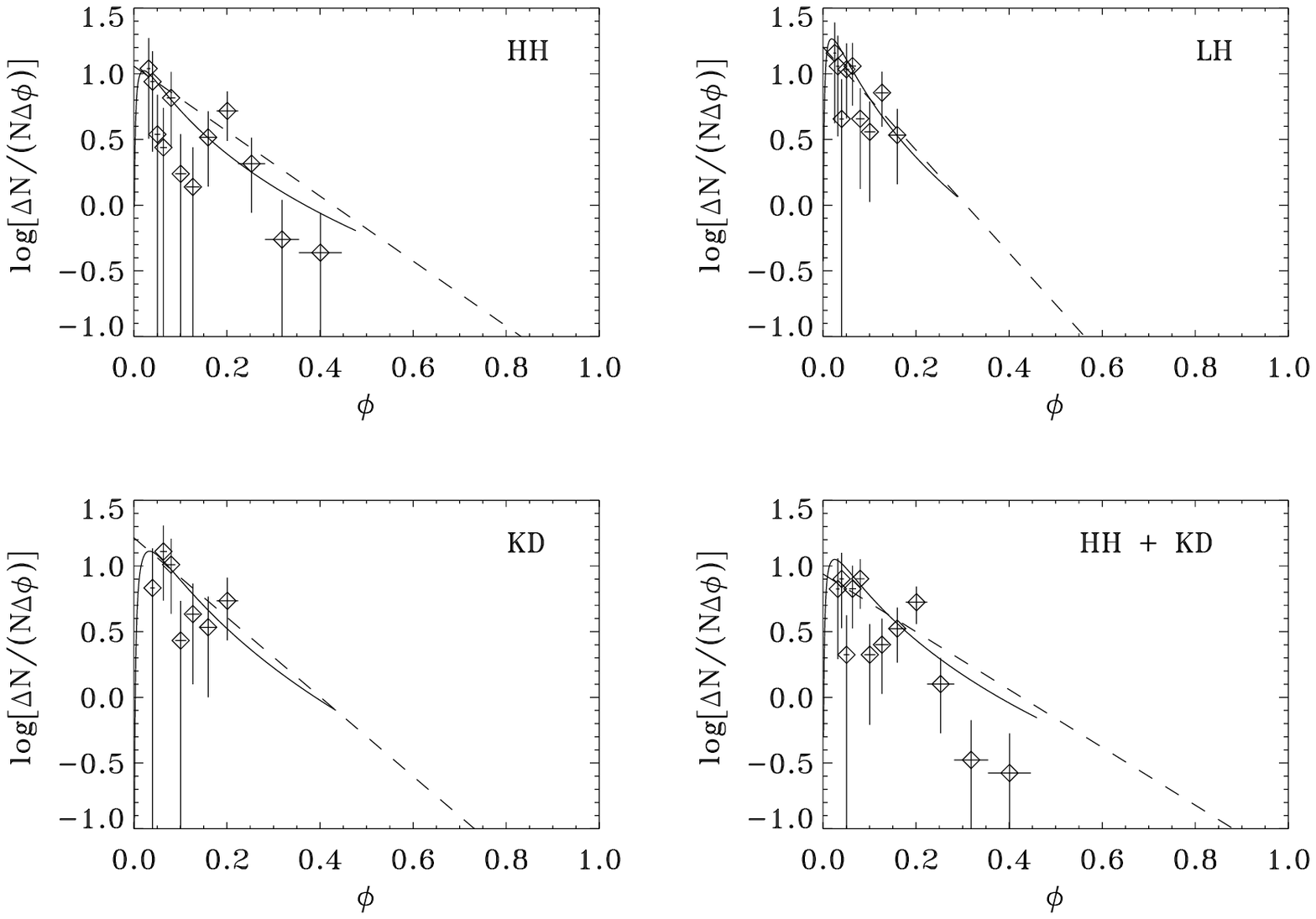}                      
\caption[ddbb]{As in Fig.\,\ref{f:Ohlk4d}, but concerning iron instead of
oxygen.}
\label{f:Felhk4d}     
\end{center}       
\end{figure}

\begin{figure}[t]  
\begin{center}      
\includegraphics[scale=0.8]{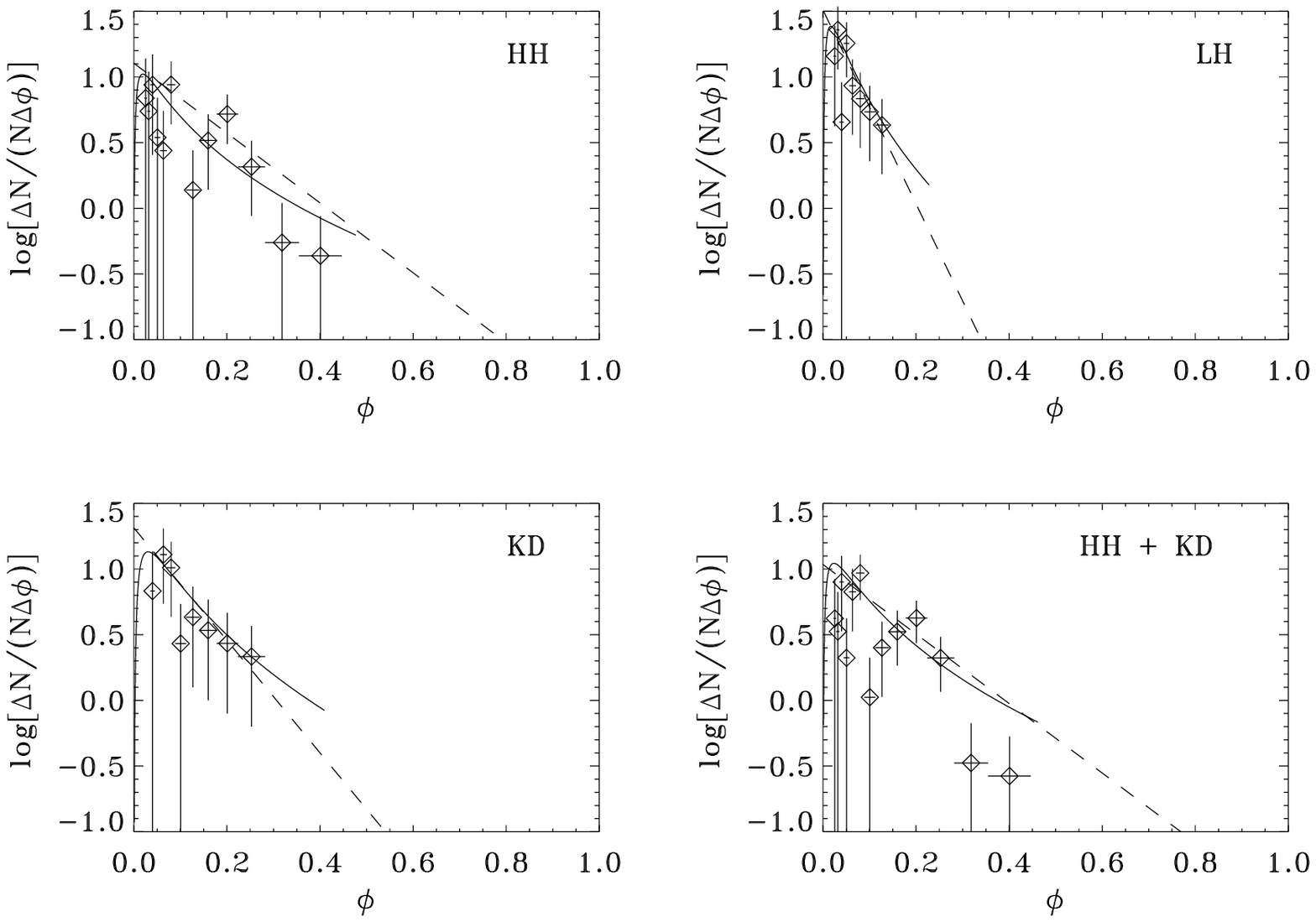}                      
\caption[ddbb]{As in Fig.\,\ref{f:Ohlk4d}, but concerning nikel instead of
oxygen.}
\label{f:Nilhk4d}     
\end{center}       
\end{figure}

The empirical differential abundance distribution,
$\psi=\Delta N/(N\Delta\phi)$, inferred from HH, LH, KD, HA
subsamples, is plotted in Figs.\,\ref{f:Ohlk4d}-\ref{f:Nilhk4d} for Q = O,
Na, Mg, Si, Ca, Ti, Cr, Fe, Ni, respectively.
Data are equally binned in [Q/H] taking $\Delta{\rm[Q/H]}=1$ dex.
Uncertainties in $\psi$, $\Delta^\mp\psi$, are calculated as Poissonian
errors, which implies $\Delta^-\psi\to\infty$ for bins populated by a single
star, $\Delta N=1$.   For further details, an interested reader is addressed
to earlier attempts (e.g., Caimmi 2011a, 2012a).

The theoretical differential abundance distribution, predicted by simple MCBR
models, is a straight line (e.g., Caimmi 2011a, 2012a) expressed as:
\begin{equation}
\label{eq:psit}
\psi=\log\left(\frac{\diff N}{N\diff\phi}\right)=
\alpha_{\rm Q}\phi+\beta_{\rm Q}~~;
\end{equation}
with regard to a selected element, Q.

Keeping in mind errors in $\psi$ are dominating on errors in $\phi$, as shown
in
Figs.\,\ref{f:Ohlk4d}-\ref{f:Nilhk4d}, regression lines have been determined
using standard least square methods (e.g., Isobe et al. 1990; Caimmi 2011b,
2012b), leaving aside points related to bins populated by a single star, where 
$\Delta^-\psi\to\infty$.   The regression procedure has been
performed on LH, HH, KD, HA subsamples and the results are shown in Table
\ref{t:rettd}.   The main features are listed below.
\begin{table*}
\caption{Regression line slope estimator, $\hat\alpha_{\rm Q}$, square root of
variance estimator, $\hat\sigma_{\hat\alpha_{\rm Q}}$, regression line 
intercept estimator, $\hat\beta_{\rm Q}$, square root of variance estimator,
$\hat\sigma_{\hat\beta_{\rm Q}}$,
for Q = O, Na, Mg,
Si, Ca, Ti, Cr, Fe, Ni, with regard to different subsamples, LH (low-$\alpha$
halo stars), HH (high-$\alpha$ halo stars), KD (low-metallicity thick
disk stars), HA (high-$\alpha$ halo + low-metallicity thick disk stars).  Bins
populated by a single star were not considered in performing the regression
procedure.   See text for further details.}
\label{t:rettd}
\begin{center}
\begin{tabular}{llllll} \hline
\multicolumn{1}{c}{Q} &
\multicolumn{1}{c}{$-\hat\alpha_{\rm Q}$} &
\multicolumn{1}{c}{$\hat\sigma_{\hat\alpha_{\rm Q}}$} &
\multicolumn{1}{c}{$\hat\beta_{\rm Q}$} &
\multicolumn{1}{c}{$\hat\sigma_{\hat\beta_{\rm Q}}$} &
\multicolumn{1}{c}{pop} \\
\hline
O  & 1.2395825D$-$00 & 4.0782960D$-$01 & 7.6627742D$-$01 & 8.4978932D$-$02 & LH \\
   & 4.7980024D$-$01 & 1.6648632D$-$01 & 4.0664611D$-$01 & 8.6555380D$-$02 & HH \\
   & 1.5989830D$-$00 & 1.6397793D$-$01 & 9.1710570D$-$01 & 6.9667135D$-$02 & KD \\
   & 4.7090995D$-$01 & 1.8054961D$-$01 & 3.1723663D$-$01 & 8.8226620D$-$02 & HA \\
Na & 7.7450493D$-$00 & 2.5744210D$-$00 & 1.6707642D$-$00 & 1.4708321D$-$01 & LH \\
   & 2.1093278D$-$00 & 7.3885505D$-$01 & 9.9040108D$-$01 & 1.4996766D$-$01 & HH \\
   & 2.9245562D$-$00 & 5.2163677D$-$01 & 1.2998827D$-$00 & 9.4153768D$-$02 & KD \\
   & 1.9183122D$-$00 & 5.1304826D$-$01 & 8.7072663D$-$01 & 9.6271620D$-$02 & HA \\
Mg & 4.0671532D$-$00 & 1.0343054D$-$00 & 1.3231002D$-$00 & 1.3072240D$-$01 & LH \\
   & 1.1714832D$-$00 & 2.8303357D$-$01 & 6.5642583D$-$01 & 1.0927166D$-$01 & HH \\
   & 2.3635566D$-$00 & 3.5784751D$-$01 & 1.1661197D$-$00 & 1.1649145D$-$01 & KD \\
   & 1.1487217D$-$00 & 4.2325854D$-$01 & 6.3913599D$-$01 & 1.4635049D$-$01 & HA \\
Si & 3.8998171D$-$00 & 8.1562678D$-$01 & 1.2767069D$-$00 & 1.0525175D$-$01 & LH \\
   & 7.7173538D$-$01 & 3.2248529D$-$01 & 5.3136082D$-$01 & 1.0414279D$-$01 & HH \\
   & 1.5024742D$-$00 & 4.5603650D$-$01 & 8.4982713D$-$01 & 1.2099418D$-$01 & KD \\
   & 7.7174327D$-$01 & 3.2248262D$-$01 & 3.1651458D$-$01 & 1.0414213D$-$01 & HA \\
Ca & 1.8159479D$-$00 & 5.1939008D$-$01 & 1.0003550D$-$00 & 8.4484996D$-$02 & LH \\
   & 7.3263802D$-$01 & 3.6421500D$-$01 & 5.2426101D$-$01 & 1.2156292D$-$01 & HH \\
   & 1.9471117D$-$00 & 8.5169618D$-$01 & 1.1180379D$-$01 & 2.1662050D$-$01 & KD \\
   & 1.9437007D$-$00 & 9.4067715D$-$01 & 7.6334270D$-$01 & 2.1532687D$-$01 & HA \\
Ti & 2.6360692D$-$00 & 1.1883298D$-$00 & 1.0953549D$-$00 & 1.5262722D$-$01 & LH \\
   & 1.3332657D$-$00 & 3.7333476D$-$01 & 7.0231670D$-$01 & 1.1503626D$-$01 & HH \\
   & 3.0193634D$-$00 & 8.9886183D$-$01 & 1.4421308D$-$00 & 2.3278547D$-$01 & KD \\
   & 1.8605943D$-$00 & 6.6369903D$-$01 & 8.5576686D$-$01 & 1.8685473D$-$01 & HA \\
Cr & 5.4302264D$-$00 & 8.2726991D$-$01 & 1.3218599D$-$00 & 7.5469008D$-$02 & LH \\
   & 2.7808727D$-$00 & 6.9731467D$-$01 & 1.0312708D$-$00 & 1.3312629D$-$01 & HH \\
   & 4.5413461D$-$00 & 1.3250681D$-$00 & 1.4005244D$-$00 & 1.7536734D$-$01 & KD \\
   & 3.9361698D$-$00 & 8.3118178D$-$01 & 1.0738609D$-$00 & 1.3953063D$-$01 & HA \\
Fe & 3.9164513D$-$00 & 1.2277461D$-$00 & 1.2025410D$-$00 & 1.1751637D$-$01 & LH \\
   & 2.4722723D$-$00 & 7.3136235D$-$01 & 1.0575561D$-$00 & 1.2057248D$-$01 & HH \\
   & 3.0259847D$-$00 & 1.4404427D$-$00 & 1.2143858D$-$00 & 2.0069136D$-$01 & KD \\
   & 2.1992895D$-$00 & 1.0676311D$-$00 & 9.3872327D$-$01 & 1.5758267D$-$01 & HA \\
Ni & 7.3799819D$-$00 & 1.3454868D$-$00 & 1.5095290D$-$00 & 1.0090129D$-$01 & LH \\
   & 2.6592341D$-$00 & 9.6597500D$-$01 & 1.1044067D$-$00 & 1.6598601D$-$01 & HH \\
   & 4.2832478D$-$00 & 6.2974963D$-$01 & 1.3132038D$-$00 & 9.6973684D$-$02 & KD \\
   & 2.6455002D$-$00 & 8.3327832D$-$01 & 1.0344389D$-$00 & 1.3131387D$-$01 & HA \\
\hline
\end{tabular}                     
\end{center}                      
\end{table*}                       
\begin{description}
\item[(1)\hspace{2.0mm}]
Regression line slope estimators, $-\hat\alpha_{\rm Q}$, are systematically
lower for HH population with respect to LH and KD population even if, in some
cases, they agree within $\mp\hat\sigma_{\hat\alpha_{\rm Q}}$.
\item[(2)\hspace{2.0mm}]
For a fixed element, regression line slope estimators may be consistent
within $\mp\hat\sigma_{\hat\alpha_{\rm Q}}$ for two populations at most, among
LH, HH, KD.
\item[(3)\hspace{2.0mm}]
Regression line intercept estimators, $\hat\beta_{\rm Q}$, are systematically
lower for HH population with respect to both LH and KD population even if,
in some cases, they agree within $\mp\hat\sigma_{\hat\beta_{\rm Q}}$.
\end{description}
In conclusion, empirical differential abundance distributions plotted in
Figs.\,\ref{f:Ohlk4d}-\ref{f:Nilhk4d} show a linear trend, as expressed by
Eq.\,(\ref{eq:psit}), leaving aside bins populated by a single star.

Arithmetic mean and rms error can be inferred from the above mentioned
distributions as:
\begin{lefteqnarray}
\label{eq:lgfi}
&& \overline{\log\phi_{\rm Q}}=\overline{\rm[Q/H]}=\frac1N\sum_{i=1}^N
{\rm[Q/H]}_i~~; \\
\label{eq:slfi}
&& \sigma_{\log\phi_{\rm Q}}=\sigma_{\rm[Q/H]}=\left\{\frac1{N-1}\sum_{i=1}^N
\left({\rm[Q/H]}_i-\overline{\rm[Q/H]}\right)^2\right\}^{1/2}~~;
\end{lefteqnarray}
where $N$ is the sample population.   Related values for each element, Q, and
subsample, LH, HH, KD, HA, are listed in Table \ref{t:parga}.
\begin{table*}
\caption{Star number, $N$, mean abundance, $\overline{[{\rm Q/H}]}$, rms error,
$\sigma_{[{\rm Q/H}]}$, Q = O, Na, Mg,
Si, Ca, Ti, Cr, Fe, Ni, inferred for different subsamples, LH (low-$\alpha$
halo stars), HH (high-$\alpha$ halo stars), KD (low-metallicity thick
disk stars), HA (high-$\alpha$ halo + low-metallicity thick disk stars).   See 
text for further details.}
\label{t:parga}
\begin{center}
\begin{tabular}{lllll} \hline
\multicolumn{1}{c}{Q} &
\multicolumn{1}{c}{$N$} &
\multicolumn{1}{c}{$\overline{[{\rm Q/H}]}$} &
\multicolumn{1}{c}{$\sigma_{[{\rm Q/H}]}$} &
\multicolumn{1}{c}{pop} \\
\hline
O  & 24 & $-$0.7517 & 0.2260 & LH \\
   & 25 & $-$0.4020 & 0.3011 & HH \\
   & 16 & $-$0.4394 & 0.2007 & KD \\
   & 41 & $-$0.4166 & 0.2643 & HA \\
Na & 24 & $-$1.3542 & 0.2595 & LH \\
   & 25 & $-$0.8760 & 0.4054 & HH \\
   & 16 & $-$1.4981 & 3.1286 & KD \\
   & 41 & $-$0.8798 & 0.3526 & HA \\
Mg & 24 & $-$0.9750 & 0.2120 & LH \\
   & 25 & $-$0.5820 & 0.3111 & HH \\
   & 16 & $-$0.6225 & 0.2191 & KD \\
   & 41 & $-$0.5978 & 0.2766 & HA \\
Si & 24 & $-$0.9713 & 0.2077 & LH \\
   & 25 & $-$0.6156 & 0.2926 & HH \\
   & 16 & $-$0.6575 & 0.2059 & KD \\
   & 41 & $-$0.6320 & 0.2602 & HA \\
Ca & 24 & $-$0.8754 & 0.2111 & LH \\
   & 25 & $-$0.6124 & 0.2807 & HH \\
   & 16 & $-$0.6813 & 0.2056 & KD \\
   & 41 & $-$0.6393 & 0.2535 & HA \\
Ti & 24 & $-$0.9638 & 0.1969 & LH \\
   & 25 & $-$0.6636 & 0.3013 & HH \\
   & 16 & $-$0.7200 & 0.2141 & KD \\
   & 41 & $-$0.6856 & 0.2692 & HA \\
Cr & 24 & $-$1.1604 & 0.2627 & LH \\
   & 25 & $-$0.9044 & 0.3451 & HH \\
   & 16 & $-$0.9663 & 0.2373 & KD \\
   & 41 & $-$0.9285 & 0.3058 & HA \\
Fe & 24 & $-$1.1404 & 0.2557 & LH \\
   & 25 & $-$0.9228 & 0.3264 & HH \\
   & 16 & $-$0.9638 & 0.2215 & KD \\
   & 41 & $-$0.9388 & 0.2877 & HA \\
Ni & 24 & $-$1.2433 & 0.2298 & LH \\
   & 25 & $-$0.9228 & 0.3435 & HH \\
   & 16 & $-$0.9588 & 0.2288 & KD \\
   & 41 & $-$0.9368 & 0.3012 & HA \\
\hline
\end{tabular}                     
\end{center}                      
\end{table*}                       

\section{Discussion} \label{disc}

While number abundances, [Q$_1/$Q$_2$], can be inferred from observations,
predictions from chemical evolution models concern mass abundances,
$Z_{\rm Q}=m_{\rm Q}/m$,
where $m_{\rm Q}$ is the total mass in the element, Q, and $m=\sum m_{\rm Q}$
is the total mass.

The normalized mass abundance, $\phi_{\rm Q}$, and the number abundance,
[Q/H], may be related as (e.g., Caimmi 2007):
\begin{lefteqnarray}
\label{eq:fiQH}
&& \log\frac{\phi_{\rm Q}}{\phi_{\rm H}}={\rm[Q/H]}~~; \\
\label{eq:fQfH}
&& \phi_{\rm Q}=\frac{Z_{\rm Q}}{(Z_{\rm Q})_\odot}~~;\qquad\phi_{\rm H}=\frac
{X}{X_\odot}~~;
\end{lefteqnarray}
where $X=Z_{\rm H}$ according to the standard notation.

The substitution of Eq.\,(\ref{eq:fiQH}) into the linear fit to the data,
Eq.\,(\ref{eq:QHOH}), yields after some algebra:
\begin{lefteqnarray}
\label{eq:fQH}
&& \frac{\phi_{\rm Q}}{\phi_{\rm H}}=\exp_{10}(b_{\rm Q})\left(\frac
{\phi_{\rm O}}{\phi_{\rm H}}\right)^{a_{\rm Q}}~~;
\end{lefteqnarray}
which, in terms of mass abundances, via Eq.\,(\ref{eq:fQfH}) translates into:
\begin{lefteqnarray}
\label{eq:ZQO}
&& Z_{\rm Q}=C_{\rm Q}(Z_{\rm O})^{a_{\rm Q}}~~; \\
\label{eq:CQ}
&& C_{\rm Q}=\exp_{10}(b_{\rm Q})\frac{(Z_{\rm Q})_\odot}{[(Z_{\rm O})_\odot]^
{a_{\rm Q}}}\left(\frac X{X_\odot}\right)^{1-a_{\rm Q}}~~;
\end{lefteqnarray}
where the dependence on $X$ may be neglected for $a_{\rm Q}$ sufficiently
close to unity and/or $X$ sufficiently close to $X_\odot$.
Accordingly, the coefficient, $C_{\rm Q}=Z_{\rm Q}/(Z_{\rm O})^{a_{\rm Q}}$,
may be conceived as a fractional generalized yield.   Related values, inferred
from the data using recent determinations of solar abundances and isotopic
fractions (Asplund et al. 2009), are listed in Table \ref{t:rette}.  A
formal calculation of solar photospheric mass abundances is shown in
Appendix \ref{a:sola}.

The special case, $a_{\rm Q}=1$, implies a linear relation between $Z_{\rm Q}$
and $Z_{\rm O}$.   Accordingly, Q and O are simple primary elements.
Conversely, $a_{\rm Q}$ different from unity outside (arbitrarily chosen)
$\mp2\sigma_{a_{\rm Q}}$ implies non-simple primary elements (i.e. appreciably
synthesised outside SNII progenitors or in absence of universal stellar
initial mass function) or secondary elements.

With regard to simple chemical evolution models, 
the assumption of instantaneous recycling and universal
stellar initial mass function implies fiducial predictions for simple primary
elements.  The special case of simple MCBR models reads (Caimmi 2011a):
\begin{lefteqnarray}
\label{eq:phiQ}
&& \phi_{\rm Q}-(\phi_{\rm Q})_i=-\frac{\hat{p}_{\rm Q}}{(1+\kappa)
(Z_{\rm Q})_\odot}\ln\frac\mu{\mu_i}~~; \\
\label{eq:phiO}
&& \phi_{\rm O}-(\phi_{\rm O})_i=-\frac{\hat{p}_{\rm O}}{(1+\kappa)
(Z_{\rm O})_\odot}\ln\frac\mu{\mu_i}~~;
\end{lefteqnarray}
where $\kappa$ is the flow parameter, positive for outflow and negative for
inflow, $\mu$ is the fractional active (i.e. viable for star formation) gas
mass normalized to the initial mass, and the index, i, denotes values at the
starting configuration.    Accordingly, the fractional yield,
$\hat{p}_{\rm Q}/\hat{p}_{\rm O}$, can be expressed as:
\begin{lefteqnarray}
\label{eq:fy}
&& \frac{\hat{p}_{\rm Q}}{\hat{p}_{\rm O}}=
\frac{Z_{\rm Q}[1-(Z_{\rm Q})_i/Z_{\rm Q}]}
{Z_{\rm O}[1-(Z_{\rm O})_i/Z_{\rm O}]}=\frac{Z_{\rm Q}}{Z_{\rm O}}~~;
\end{lefteqnarray}
which is owing to a further assumtion of MCBR models, that all elements are
simple primary i.e. constant ratio, $Z_{\rm Q}/Z$, $Z=\sum Z_{\rm Q}$ (Caimmi
2011a) implying, in turn, constant ratio, $Z_{\rm Q}/Z_{\rm O}$.   A formal
calculation is shown in Appendix \ref{a:fry}.

The substitution of Eq.\,(\ref{eq:fy}) into (\ref{eq:fQH}), the last
particularized to the unit slope, produces:
\begin{lefteqnarray}
\label{eq:fg}
&& \frac{\hat{p}_{\rm Q}}{\hat{p}_{\rm O}}=\frac{(Z_{\rm Q})_\odot}
{(Z_{\rm O})_\odot}\exp_{10}(b_{\rm Q})~~;
\end{lefteqnarray}
where the intercepts, $b_{\rm Q}$, are listed in Table\,\ref{t:rette}.
In conclusion, simple MCBR chemical evolution models imply $a_{\rm Q}=1$.

Let a generic element, Q $\ne$ O, be considered as simple primary if the
regression line slope estimator, inferred from the empirical [Q/H]-[O/H]
relation, is consistent with the unit slope within
$\mp2\hat\sigma_{\hat a_{\rm Q}}$.   With regard
to the subsamples studied in the current attempt,
an inspection of Table \ref{t:rette} shows the following elements are inferred
from the data to be simple primary.   LH: Ni within
$\mp1\hat\sigma_{\hat a_{\rm Q}}$
and Na, Mg, Si, Ca, Ti, within $\mp2\hat\sigma_{\hat a_{\rm Q}}$, while Cr and
Fe are excluded.   HH: Mg, Si, Ti, within $\mp1\hat\sigma_{\hat a_{\rm Q}}$
and Ca within about $\mp2\hat\sigma_{\hat a_{\rm Q}}$, while Na, Cr, Fe, Ni, 
are excluded. KD: Si, Ca, Ti, within $\mp1\hat\sigma_{\hat a_{\rm Q}}$ and Mg,
Fe, Ni, within $\mp2\hat\sigma_{\hat a_{\rm Q}}$, while Na and Cr are
excluded.   Then
$\alpha$ elements (Mg, Si, Ca, Ti) together with Na, Ni, for LH stars
and Fe, Ni, for KD stars, are inferred from the data to be simple primary
elements, which does not hold for Na (HH and KD stars), Cr (LH, HH and KD
stars), Fe (LH and HH stars), Ni (HH stars).

Keeping in mind $Z_{\rm Q}\ll1$, an exponent, $a_{\rm Q}>1$, appearing in
Eq.\,(\ref{eq:ZQO}), implies $Z_{\rm Q}$ grows at an increasing rate with
respect to $Z_{\rm O}$, as expected for non-simple primary or secondary
elements.   In this view, Na, Cr, Fe, Ni, could be conceived as non-simple
primary or secondary elements, which implies [O/Q] is decreasing in time, as
shown by the data (e.g., Ra12, Fig.\,8 therein).

An empirical [Na/H]-[Fe/H] relation has been inferred from a large $(N=1891)$
sample in a recent attempt (Carretta 2013).   Aiming to a comparison with the
current results, the particularization of Eq.\,(\ref{eq:QHOH}) to Q = Na, Fe,
after elimination of [O/H] yields:
\begin{lefteqnarray}
\label{eq:NaFe}
&& {\rm[Na/H]}=A{\rm[Fe/H]}+B~~; \\
\label{eq:AB}
&& A=\frac{a_{\rm Na}}{a_{\rm Fe}}~~;\qquad
B=b_{\rm Na}-Ab_{\rm Fe}~~;
\end{lefteqnarray}
where the values of the coefficients on the right-hand side are listed in
Table\,\ref{t:rette} for different subsamples.   Accordingly, $A$ and $B$ can
be evaluated for the ``main sequences'' on the
$\{{\sf O}{\rm [O/H]}{\rm[Na/H]}\}$ and $\{{\sf O}{\rm [O/H]}{\rm[Fe/H]}\}$
plane described in a recent attempt (Caimmi 2013)%
\footnote
{The main sequence in the caption of Fig.\,2 therein is expressed as
${\rm[Na/H]}={\rm[Fe/H]}-0.4\mp0.3$ instead of
${\rm[Na/H]}=1.25{\rm[Fe/H]}-0.4\mp0.3$, due to a printing error.}.   %
Starting from $a_{\rm Fe}=1$, $b_{\rm Fe}=-0.45, -0.70, -0.20$;
$a_{\rm Na}=1.25$, $b_{\rm Na}=-0.40, -0.70, -0.10$; (Caimmi 2013), the
result is $A=1.25$, $B=0.1625, -0.4000, 0.7250$.   For subsamples considered
in the current paper, $A$ and $B$ can be directly evaluated by determining the
regression line on the $\{{\sf O}{\rm[Fe/H]}{\rm[Na/H]}\}$ plane.
The results are listed in Table\,\ref{t:FeNa}.
\begin{table*}
\caption{Regression line slope estimator, $\hat A$, square root of
variance estimator, $\hat\sigma_{\hat A}$, regression line 
intercept estimator, $\hat B$, square root of variance estimator,
$\hat\sigma_{\hat B}$, inferred from the data with regard to different 
subsamples, LH (low-$\alpha$
halo stars), HH (high-$\alpha$ halo stars), KD (low-metallicity thick
disk stars), HA (high-$\alpha$ halo + low-metallicity thick disk stars),
and the total sample with the exclusion of OL stars, HK = LH+HH+KD.  See
text for further details.}
\label{t:FeNa}
\begin{center}
\begin{tabular}{lllll} \hline
\multicolumn{1}{c}{$\hat A$} &
\multicolumn{1}{c}{$\hat\sigma_{\hat A}$} &
\multicolumn{1}{c}{$\hat B$} &
\multicolumn{1}{c}{$\hat\sigma_{\hat B}$} &
\multicolumn{1}{c}{pop} \\
\hline
1.0147770D$-$00 & 8.2903367D$-$02 & $-$1.9689805D$-$01 & 1.0642713D$-$01 & LH \\
1.2409084D$-$00 & 3.5227278D$-$02 & $+$2.6712483D$-$01 & 3.1356340D$-$02 & HH \\
1.1821507D$-$00 & 7.6558086D$-$02 & $+$2.5367275D$-$01 & 6.2117800D$-$02 & KD \\
1.2248702D$-$00 & 3.7797350D$-$02 & $+$2.6893320D$-$01 & 3.2037882D$-$02 & HA \\
1.3043300D$-$00 & 5.5544051D$-$02 & $+$2.1084026D$-$01 & 6.4229880D$-$02 & HK \\
\hline
\end{tabular}                     
\end{center}                      
\end{table*}                       

Related [Na/H]-[Fe/H] relations can be plotted as straight lines and compared
with their counterparts inferred from the large sample (Carretta 2013).
Unfortunately, the regression line is not expressed therein and the comparison
has to be made by eye.   Interestingly, outliers specified within the large
sample (Carretta 2013) could be related to LH population.   An inspection of
Table\,\ref{t:FeNa} shows a lower slope for LH subsample with respect to the
other ones and, in fact, outliers exhibit lower slope with respect to
``normal'' stars (Carretta 2013, Fig.\,5 therein, bottom panel).   In
addition, it can be seen most stars belonging to the large sample lie within
the main sequence, [Na/H] = 1.25\,[Fe/H]+0.1625$\mp0.5625$, with the
exception of a few Na-overabundant, low-metallicity stars which, to this
respect, should be considered as ``outliers'' instead of Na-deficient,
low-metallicity stars.

Turning to the whole set of elements considered in the current attempt, it
would be relevant investigating to what extent simple MCBR
models fit to the data.   With regard to a selected element, Q, the
slope of the theoretical differential abundance distribution, expressed by
Eq.\,(\ref{eq:psit}), can be explicitly written as (e.g., Caimmi 2011a,
2012a):
\begin{lefteqnarray}
\label{eq:alfaQ}
&& \alpha_{\rm Q}=-\frac{1+\kappa}{\ln10}\frac{(Z_{\rm Q})_\odot}
{\hat{p}_{\rm Q}}~~;
\end{lefteqnarray}
and the ratio of the terms on both sides of Eq.\,(\ref{eq:alfaQ}) to their
counterparts particularized to oxygen, Q = O, after little algebra yields:
\begin{lefteqnarray}
\label{eq:alOQ}
&& \frac{\hat{p}_{\rm Q}}{\hat{p}_{\rm O}}=\frac{(Z_{\rm Q})_\odot}
{(Z_{\rm O})_\odot}\frac{\alpha_{\rm O}}{\alpha_{\rm Q}}~~;
\end{lefteqnarray}
to be compared with Eq.\,(\ref{eq:fg}).   Related rms errors are expressed in
Appendix \ref{a:err}, Eqs.\,(\ref{eq:sgQOa}) and (\ref{eq:sgQOb}),
respectively.  The results are shown in Table \ref{t:fracy} for Q = Na, Mg,
Si, Ca, Ti, Cr, Fe, Ni, with regard to LH, HH, KD, HA subsamples.  An
inspection of Table \ref{t:fracy} discloses the following.
\begin{table*}
\caption{Fractional yield, $\hat{p}_{\rm Q}/\hat{p}_{\rm O}$,
inferred from the data in the light of simple MCBR models via
Eqs.\,(\ref{eq:fg}), (\ref{eq:alOQ}), related rms error,
$\hat\sigma_{\hat{p}_{\rm Q}/\hat{p}_{\rm O}}$, inferred from
Eqs.\,(\ref{eq:sgQOb}), (\ref{eq:sgQOa}), respectively,
intercept of the straight line, $b_{\rm Q}$, expressed by
Eq.\,(\ref{eq:QHOH}), inferred from the data in the light of simple MCBR
models via Eq.\,(\ref{eq:bQ}), related rms error inferred from 
Eq.\,(\ref{eq:sgbQ}), for Q = Na, Mg,
Si, Ca, Ti, Cr, Fe, Ni, with regard to different subsamples, LH (low-$\alpha$
halo stars), HH (high-$\alpha$ halo stars), KD (low-metallicity thick
disk stars), HA (high-$\alpha$ halo + low-metallicity thick disk stars).   
See text for further details.}
\label{t:fracy}
\begin{center}
\begin{tabular}{llllllll} \hline
\multicolumn{1}{c}{Q} &
\multicolumn{1}{c}{$\hat{p}_{\rm Q}/\hat{p}_{\rm O}$} &
\multicolumn{1}{c}{$\sigma_{\hat{p}_{\rm Q}/\hat{p}_{\rm O}}$} &
\multicolumn{1}{c}{$\hat{p}_{\rm Q}/\hat{p}_{\rm O}$} &
\multicolumn{1}{c}{$\sigma_{\hat{p}_{\rm Q}/\hat{p}_{\rm O}}$} &
\multicolumn{1}{c}{$-b_{\rm Q}$} &
\multicolumn{1}{c}{$\sigma_{b_{\rm Q}}$} &
\multicolumn{1}{c}{pop} \\
\hline
Na & 1.6415D$-$03 & 1.3709D$-$04 & 8.1601D$-$04 & 3.8164D$-$04 & +7.9575D$-$01 & 2.0311D$-$01 & LH \\
   & 2.3585D$-$03 & 6.1201D$-$05 & 1.1597D$-$03 & 3.7181D$-$04 & +6.4308D$-$01 & 2.1413D$-$01 & HH \\
   & 2.4810D$-$03 & 1.1758D$-$04 & 2.7876D$-$03 & 5.7353D$-$04 & +2.6222D$-$01 & 8.9354D$-$02 & KD \\
   & 2.4176D$-$03 & 5.2766D$-$05 & 1.2516D$-$03 & 5.8508D$-$04 & +6.0998D$-$01 & 2.0302D$-$01 & HA \\
Mg & 6.6340D$-$02 & 2.3779D$-$03 & 3.7642D$-$02 & 1.5653D$-$02 & +5.1602D$-$01 & 1.8059D$-$01 & LH \\
   & 8.4159D$-$02 & 1.4535D$-$03 & 5.0584D$-$02 & 2.1388D$-$02 & +3.8768D$-$01 & 1.8363D$-$01 & HH \\
   & 8.8889D$-$02 & 1.7901D$-$03 & 8.3553D$-$02 & 1.5279D$-$02 & +1.6972D$-$01 & 7.9417D$-$02 & KD \\
   & 8.5074D$-$02 & 1.1700D$-$03 & 5.0630D$-$02 & 2.6923D$-$02 & +3.8728D$-$01 & 2.3094D$-$01 & HA \\
Si & 6.0818D$-$02 & 1.9922D$-$03 & 3.6878D$-$02 & 1.4377D$-$02 & +4.9777D$-$01 & 1.6931D$-$01 & LH \\
   & 6.9121D$-$02 & 1.2911D$-$03 & 7.2132D$-$02 & 3.9179D$-$02 & +2.0641D$-$01 & 2.3589D$-$01 & HH \\
   & 7.2081D$-$02 & 1.3738D$-$03 & 1.2347D$-$01 & 3.9559D$-$02 & -2.7037D$-$02 & 1.3914D$-$01 & KD \\
   & 6.9611D$-$02 & 1.0363D$-$03 & 7.0795D$-$02 & 4.0148D$-$02 & +2.1453D$-$01 & 2.4629D$-$01 & HA \\
Ca & 7.5105D$-$03 & 2.3113D$-$04 & 7.6425D$-$03 & 3.3317D$-$03 & +1.6583D$-$01 & 1.8933D$-$01 & LH \\
   & 6.4771D$-$03 & 9.9101D$-$05 & 7.3322D$-$03 & 4.4451D$-$03 & +1.8383D$-$01 & 2.6329D$-$01 & HH \\
   & 6.5755D$-$03 & 1.1070D$-$04 & 9.1942D$-$03 & 4.1307D$-$03 & +8.5547D$-$02 & 1.9512D$-$01 & KD \\
   & 6.4478D$-$03 & 8.3565D$-$05 & 2.7125D$-$03 & 1.6748D$-$03 & +6.1569D$-$01 & 2.6815D$-$01 & HA \\
Ti & 2.6767D$-$04 & 8.7155D$-$06 & 2.5617D$-$04 & 1.4296D$-$04 & +3.2768D$-$01 & 2.4237D$-$01 & LH \\
   & 2.9850D$-$04 & 4.4312D$-$06 & 1.9604D$-$04 & 8.7412D$-$05 & +4.4386D$-$01 & 1.9364D$-$01 & HH \\
   & 3.0545D$-$04 & 7.5481D$-$06 & 2.8849D$-$04 & 9.0837D$-$05 & +2.7607D$-$01 & 1.3675D$-$01 & KD \\
   & 2.9843D$-$04 & 3.7205D$-$06 & 1.3788D$-$04 & 7.2204D$-$05 & +5.9671D$-$01 & 2.2743D$-$01 & HA \\
Cr & 1.4977D$-$03 & 4.4001D$-$05 & 6.6163D$-$04 & 2.3988D$-$04 & +6.4154D$-$01 & 1.5746D$-$01 & LH \\
   & 1.0435D$-$03 & 1.6988D$-$05 & 5.0007D$-$04 & 2.1409D$-$04 & +7.6312D$-$01 & 1.8593D$-$01 & HH \\
   & 1.0362D$-$03 & 2.8905D$-$05 & 1.0205D$-$03 & 3.1562D$-$04 & +4.5334D$-$01 & 1.3432D$-$01 & KD \\
   & 1.0365D$-$03 & 1.4518D$-$05 & 3.4675D$-$04 & 1.5178D$-$04 & +9.2214D$-$01 & 1.9010D$-$01 & HA \\
Fe & 1.1563D$-$01 & 3.6434D$-$03 & 7.1386D$-$02 & 3.2441D$-$02 & +4.9962D$-$01 & 1.9736D$-$01 & LH \\
   & 7.3818D$-$02 & 1.4793D$-$03 & 4.3772D$-$02 & 1.9959D$-$02 & +7.1204D$-$01 & 1.9803D$-$01 & HH \\
   & 7.4881D$-$02 & 2.0567D$-$03 & 1.1918D$-$01 & 5.8035D$-$02 & +2.7702D$-$01 & 2.1148D$-$01 & KD \\
   & 7.3989D$-$02 & 1.2047D$-$03 & 4.8293D$-$02 & 2.9874D$-$02 & +6.6934D$-$01 & 2.6865D$-$01 & HA \\
Ni & 4.1266D$-$03 & 1.5637D$-$04 & 2.0896D$-$03 & 7.8597D$-$04 & +7.7478D$-$01 & 1.6336D$-$01 & LH \\
   & 4.2712D$-$03 & 8.7387D$-$05 & 2.2446D$-$03 & 1.1276D$-$03 & +7.4370D$-$01 & 2.1817D$-$01 & HH \\
   & 4.3347D$-$03 & 1.2423D$-$04 & 4.6441D$-$03 & 8.3250D$-$04 & +4.2793D$-$01 & 7.7851D$-$02 & KD \\
   & 4.2930D$-$03 & 7.0677D$-$05 & 2.2144D$-$03 & 1.0988D$-$03 & +7.4957D$-$01 & 2.1550D$-$01 & HA \\
\hline
\end{tabular}                     
\end{center}                      
\end{table*}                       
\begin{description}
\item[(1)\hspace{2.0mm}]
For assigned element, Q, and population, the results for $\hat{p}_{\rm Q}/
\hat{p}_{\rm O}$ are consistent within $\mp\sigma_{\hat{p}_{\rm Q}/
\hat{p}_{\rm O}}$ or less, leaving aside Ca, Ti, Cr (LH, HH, HA), Ni (LH).
\item[(2)\hspace{2.0mm}]
For assigned element, Q, the results for $\hat{p}_{\rm Q}/\hat{p}_{\rm O}$
are consistent within $\mp2\sigma_{\hat{p}_{\rm Q}/\hat{p}_{\rm O}}$ or
less for HH, KD, HA populations, while the contrary holds for LH population,
which exhibits lower values with regard to Na, Mg, Si, Ti, higher values with
respect to Ca, Cr, Fe, and nearly equal values for Ni, in connection with
Eqs.\,(\ref{eq:fg}) and (\ref{eq:sgQOb}).
\item[(3)\hspace{2.0mm}]
A similar trend, partially hidden by larger errors, is shown via
Eqs.\,(\ref{eq:alOQ}) and (\ref{eq:sgQOa}).   In particular, larger
$\hat{p}_{\rm Fe}/\hat{p}_{\rm O}$ values imply a lower [O/Fe] abundance ratio
for LH population with respect to HH, KD, HA, as inferred from the data.
Accordingly, MCBR models might provide a viable description of the chemical
evolution of the halo and the (low-metallicity) thick disk.
\end{description}

\begin{figure}[t]  
\begin{center}      
\includegraphics[scale=0.8]{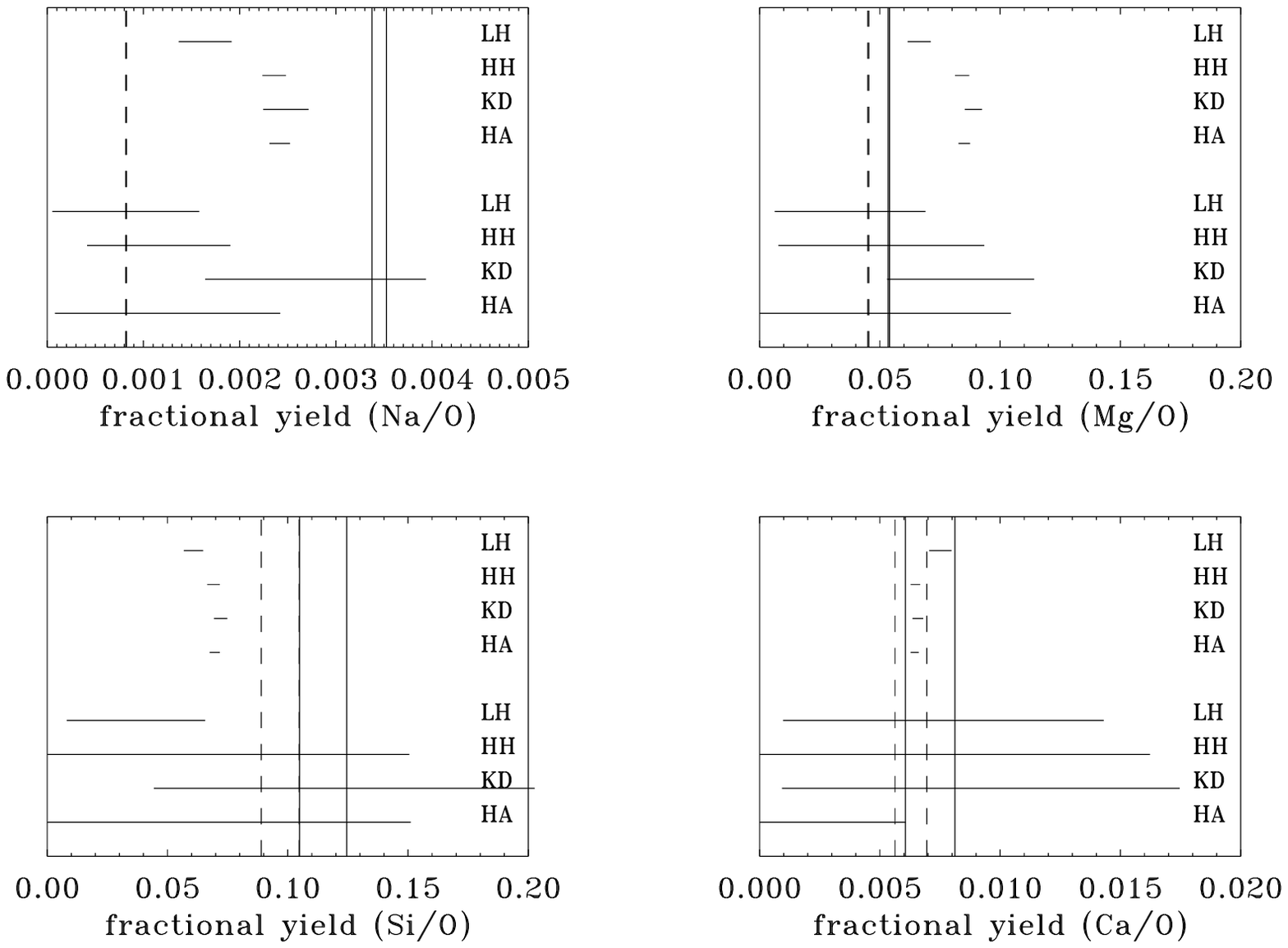}                      
\caption[ddbb]{
Comparison between the fractional yield, $\hat{p}_{\rm Q}/\hat{p}_{\rm O}$,
Q = Na, Mg, Si, Ca, inferred from Eqs.\,(\ref{eq:fg}) and (\ref{eq:alOQ}), for
different subsamples as indicated (top and bottom bars, respectively)
and theoretical counterparts deduced from stellar nucleosynthesis (vertical
bands) for solar, $Z=Z_\odot$ (full), and subsolar, $Z=Z_\odot/10$ (dashed)
metallicity, under the
assumption of a power-law stellar initial mass function.   The bar
semiamplitude equals $2\sigma_{\hat{p}_{\rm Q}/\hat{p}_{\rm O}}$.   The band
width relates to a fiducial range of power-law exponent, $-3\le-p\le-2$.   
See text for further details.}
\label{f:fryc1}     
\end{center}       
\end{figure}

\begin{figure}[t]  
\begin{center}      
\includegraphics[scale=0.8]{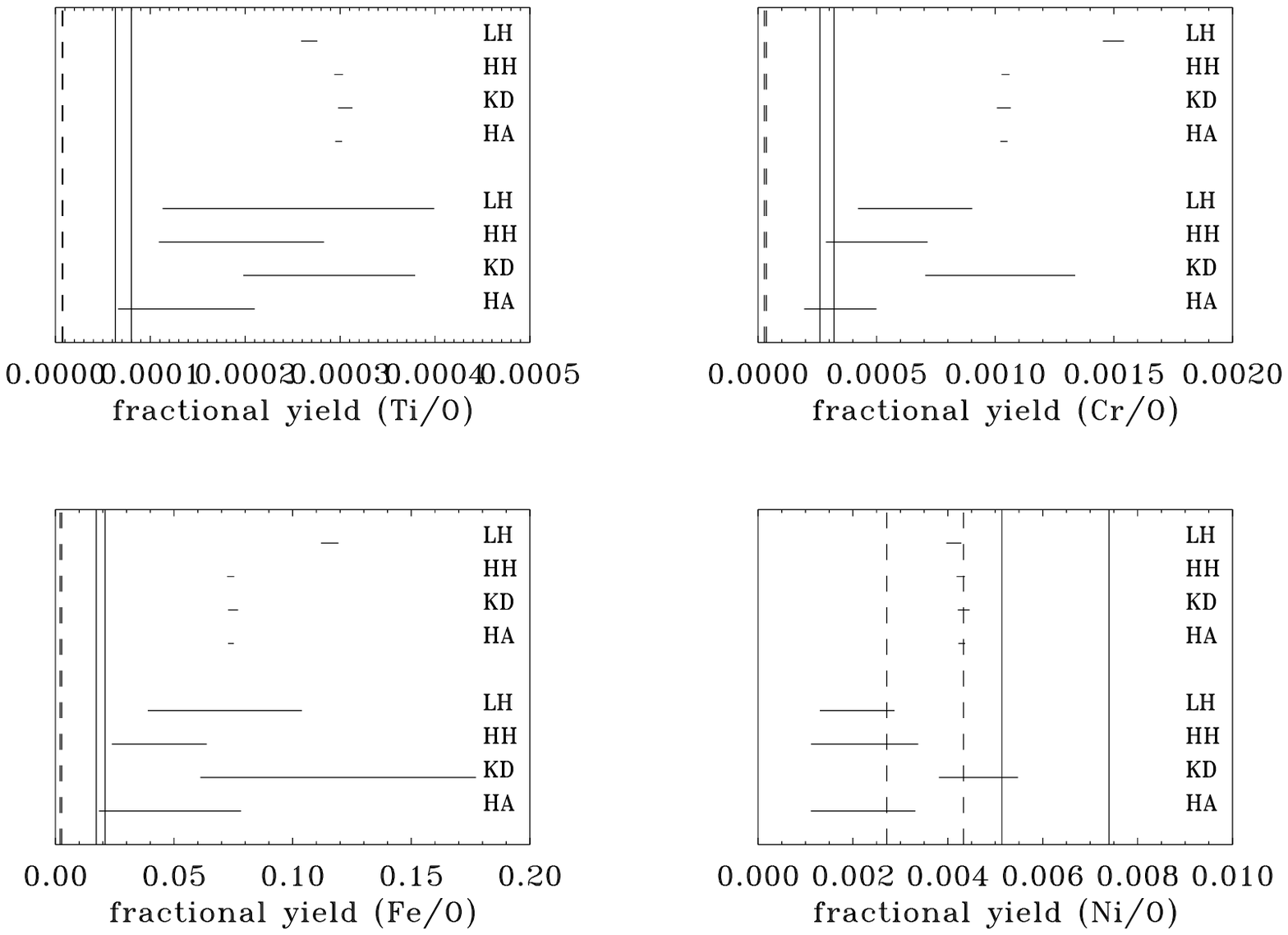}                      
\caption[ddbb]{As in Fig.\,\ref{f:fryc1}, but for Q = Ti, Cr, Fe, Ni.}
\label{f:fryc2}     
\end{center}       
\end{figure}

A comparison between the fractional yield, $\hat{p}_{\rm Q}/\hat{p}_{\rm O}$,
inferred from the data in the framework of simple MCBR models, and
theoretical counterparts deduced from an earlier attempt (Woosley and Weaver
1995) is shown in Figs.\,\ref{f:fryc1} and \ref{f:fryc2} for Q = Na, Mg, Si,
Ca, and Q = Ti, Cr, Fe, Ni, respectively, with regard to different subsamples
(LH, HH, KD, HA) and different power-law stellar initial mass function
exponents.

More specifically, horizontal bars represent fractional yields inferred from
Eqs.\,(\ref{eq:fg}) and (\ref{eq:alOQ}), top and bottom, respectively, where
the semiamplitude equals $2\sigma_{\hat{p}_{\rm Q}/\hat{p}_{\rm O}}$ in each
case, as listed in Table \ref{t:fracy}.   Full and
dashed vertical bands represent theoretical fractional yields deduced from
SNII progenitor nucleosynthesis within the mass range, $11\le m/m_\odot\le40$,
$Z=0.02$, and $12\le m/m_\odot\le40$, $Z=0.002$, respectively (Woosley
and Weaver 1995, model A), where the power-law stellar initial mass function
exponent, $p$, lies within the range, $-3\le-p\le-2$.   A narrow band implies
little dependence of fracional yields on $p$ and vice versa.
A formal
expression of the theoretical fractional yield is shown in Appendix
\ref{a:try}.
 
An inspection of Figs.\,\ref{f:fryc1} and \ref{f:fryc2} discloses that
empirical, inferred from Eqs.\,(\ref{eq:fg}) and (\ref{eq:alOQ}), and
theoretical fractional yields are consistent (in the sense that
horizontal bars related to the former lie between vertical bands related to
the latter) for Na, Ca, Ni, and Na, Mg, Ca, Ni, respectively, while the
contrary holds for the remaining elements in connection with one population at
least.   The
discrepancy could be due to a number of reasons, for instance (i) subsamples
are poorly populated and different regression lines might be related to richer
subsamples; (ii) Ti, Cr, Fe, (at least) are appreciably synthesised outside
SNII progenitors e.g., SNIa progenitors and AGB stars; (iii) updated models
make O production reduced by a factor of about 2 and Ti, Cr, Fe production
increased by a comparable factor; (iv) lower empirical fractional yields are
expected in presence of significant cosmic scatter provided it is more
efficient for light elements with respect to heavy elements.

The substitution of Eq.\,(\ref{eq:alOQ}) into (\ref{eq:fg}) produces:
\begin{lefteqnarray}
\label{eq:bQ}
&& b_{\rm Q}=\log\frac{\alpha_{\rm O}}{\alpha_{\rm Q}}~~;
\end{lefteqnarray}
which is the intercept of the straight line, expressed by
Eq.\,(\ref{eq:QHOH}), inferred from the data in the light of simple MCBR
models, implying $a_{\rm Q}=1$ via Eq.\,(\ref{eq:OQOH}).  Values of
intercept, $b_{\rm Q}$, and related rms error, $\sigma_{b_{\rm Q}}$,expressed
by Eqs.\,(\ref{eq:bQ}), (\ref{eq:sgbQ}), are listed in Table \ref{t:fracy}.
The comparison with their counterparts, listed in Table \ref{t:rette}, shows
results consistent within $\mp2\sigma_{b_{\rm Q}}$ or less, for assigned
element, Q, and subsample, LH, HH, KD, HA.

\begin{figure}[t]  
\begin{center}      
\includegraphics[scale=0.8]{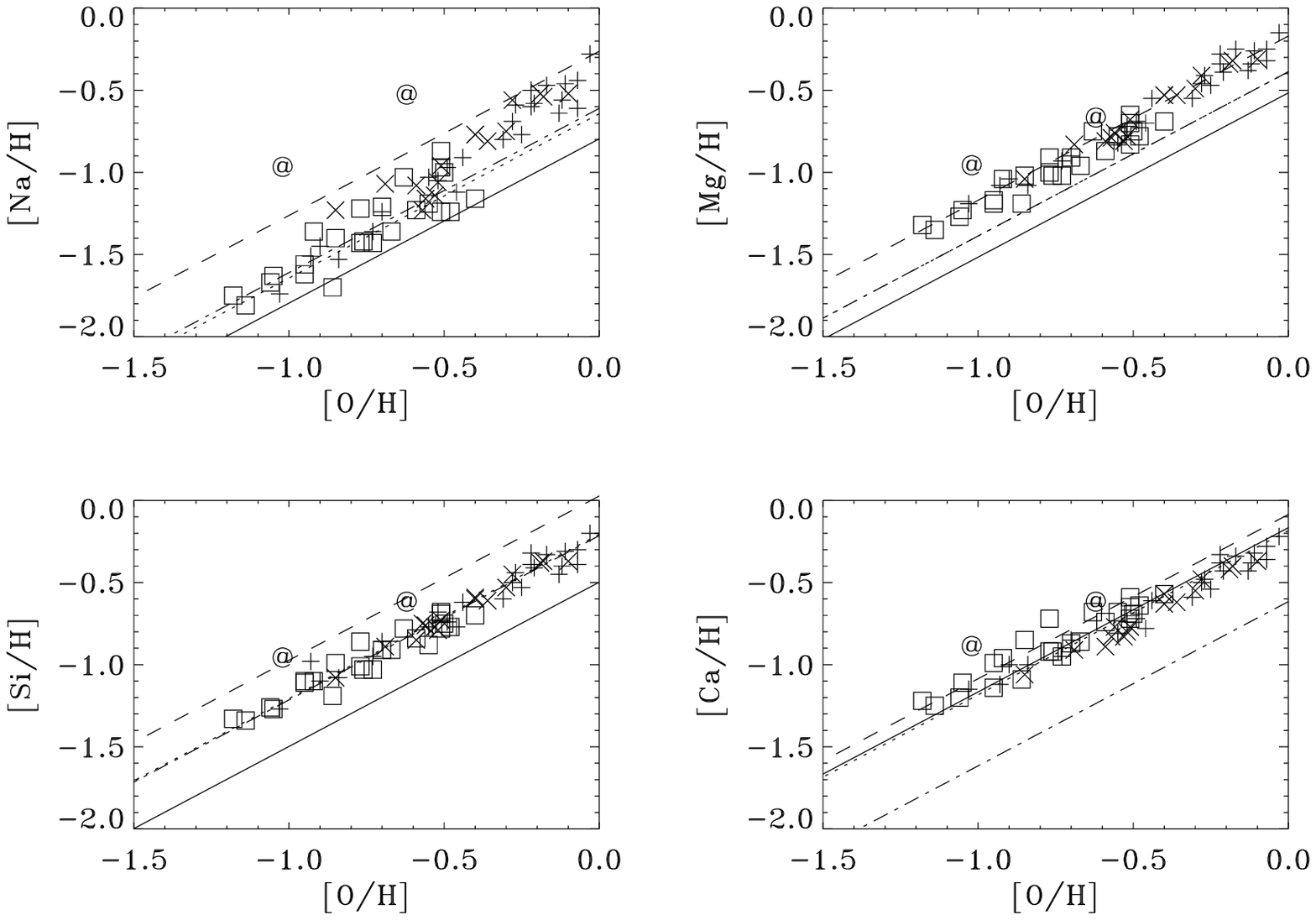}                      
\caption[ddbb]{The theoretical [Q/H]-[O/H] relation, Q = Na, Mg, Si, Ca,
inferred from simple MCBR models, via Eq.\,(\ref{eq:bQ}), for
subsamples, LH (low-$\alpha$ halo stars, full lines), HH (high-$\alpha$ halo
stars, dotted lines), KD (low-metallicity thick disk stars, dashed lines), HA
(high-$\alpha$ halo + low-metallicity thick disk stars, dot-dashed lines).   
Subsample stars are also plotted with same symbols as in
Fig.\,\ref{f:NaMgSiCa}.   See text for further details.}
\label{f:NaMgSiCa4r}     
\end{center}       
\end{figure}

\begin{figure}[t]  
\begin{center}      
\includegraphics[scale=0.8]{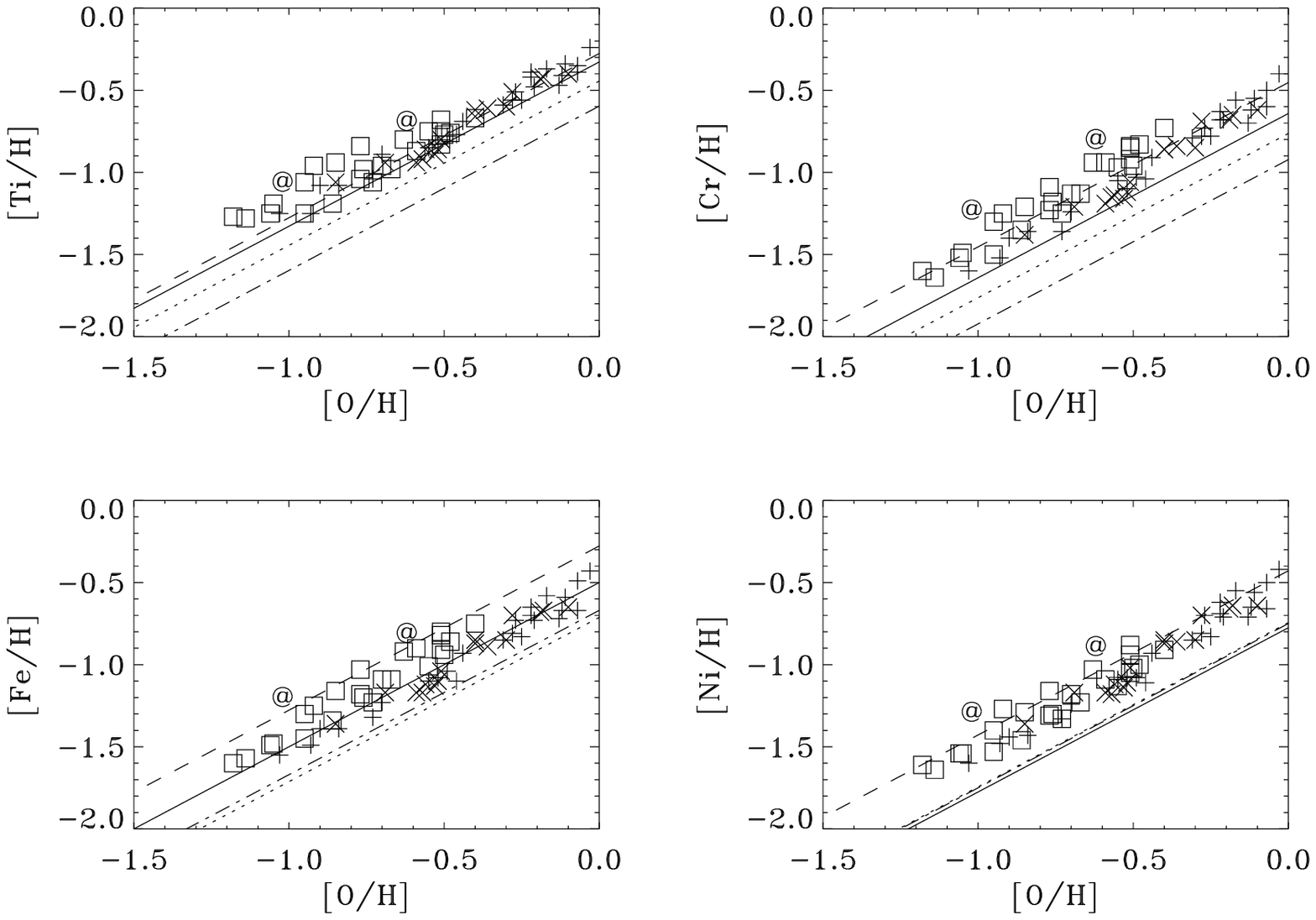}                      
\caption[ddbb]{As in Fig.\,\ref{f:NaMgSiCa4r}, but for Q = Ti, Cr, Fe, Ni.
}
\label{f:TiCrFeNi4r}     
\end{center}       
\end{figure}

The straight lines of unit slope and intercept, inferred from
Eq.\,(\ref{eq:bQ}), are plotted in Figs.\,\ref{f:NaMgSiCa4r} and 
\ref{f:TiCrFeNi4r} for Q = Na, Mg, Si, Ca, and Q = Ti, Cr, Fe, Ni,
respectively, and compared to the data.
%
The above mentioned lines are consistent with the main sequences enclosing the
data, shown in Figs.\,\ref{f:NaMgSiCa} and \ref{f:TiCrNiFe}, within
$\mp2\sigma_{b_{\rm Q}}$ or less.

The cut parameter (ratio of element abundance within the flowing gas to its
counterpart within the pre-existing gas), $\zeta_{\rm Q}$, in the case under
discussion is expressed as (Caimmi 2011a, 2012a):
\begin{lefteqnarray}
\label{eq:zitQ1}
&& \zeta_{\rm Q}=1-\frac{A_{\rm Q}\hat{p}_{\rm Q}}\kappa~~;\qquad A_{\rm Q}=
\frac{Z_\odot}{(Z_{\rm Q})_\odot}~~;
\end{lefteqnarray}
where $Z$ is the total metal abundance.   The substitution of
Eq.\,(\ref{eq:alfaQ}) into (\ref{eq:zitQ1}) after some algebra yields:
\begin{lefteqnarray}
\label{eq:zitQ2}
&& \zeta_{\rm Q}=1+\frac{Z_\odot}{\ln10}\frac1{\alpha_{\rm Q}}\frac{1+\kappa}
\kappa~~;
\end{lefteqnarray}
in the limit of strong outflow, $\kappa\gg1$, $\alpha_{\rm Q}\ll-1$, which
implies $\zeta_{\rm Q}\appleq1$, as expected.

With regard to a selected element, Q, the assumption of a universal stellar
initial mass function implies constant yield, $\hat{p}_{\rm Q}$, for different
populations, P1 and P2 say.   Accordingly, the following relation is inferred
from Eq.\,(\ref{eq:alfaQ}):
\begin{lefteqnarray}
\label{eq:ap12}
&& \frac{(\alpha_{\rm Q})_{\rm P1}}{(\alpha_{\rm Q})_{\rm P2}}=\frac
{1+\kappa_{\rm P1}}{1+\kappa_{\rm P2}}~~;
\end{lefteqnarray}
where the ratio on the right-hand side may be conceived as an indicator of the
flow parameter ratio between the populations, P1 and P2; P$i$ = LH, HH, KD,
HA, $i$ = 1, 2.   Computed values,
$(F_{\rm Q})_{\rm XY}=(\alpha_{\rm Q})_{\rm XY}/(\alpha_{\rm Q})_{\rm LH}$,
together with related rms errors,
$\sigma_{(F_{\rm Q})_{\rm XY}}=\sigma_{(\alpha_{\rm Q})_{\rm XY}/
(\alpha_{\rm Q})_{\rm LH}}$, expressed by Eq.\,(\ref{eq:sgQOb}), assumed lower
and upper limit,
$(F_{\rm Q})_{\rm XY}^\mp=(\alpha_{\rm Q})_{\rm XY}/(\alpha_{\rm Q})_{\rm LH}
\mp2\sigma_{(\alpha_{\rm Q})_{\rm XY}/(\alpha_{\rm Q})_{\rm LH}}$, for Q = O,
Na, Mg, Si, Ca, Ti, Cr, Fe, Ni; P1 = HH, KD, HA; P2 = LH; are listed in Table
\ref{t:flowr}.
\begin{table*}
\caption{Slope ratio,
$(F_{\rm Q})_{\rm XY}=(\alpha_{\rm Q})_{\rm XY}/(\alpha_{\rm Q})_{\rm LH}$,
related rms error, $\sigma_{(F_{\rm Q})_{\rm XY}}$, assumed lower and upper
limit, $(F_{\rm Q})_{\rm XY}^\mp=(F_{\rm Q})_{\rm XY}\mp2\sigma_{(F_{\rm Q})_
{\rm XY}}$, for different elements, Q = O, Na, Mg, Si, Ca, Ti, Cr, Fe, Ni, and
different subsamples, XY = HH, KD, HA.   With regard to lower limits,
$(F_{\rm Q})_{\rm XY}^-$, unphysical negative values are replaced by null
values.   The intersection of assumed validity ranges,
$(F_{\rm Q})_{\rm XY}^\mp=(F_{\rm Q})_{\rm XY}\mp2
\sigma_{(F_{\rm Q})_{\rm XY}}$, is denoted as $\cap_{\rm Q}$ and
related values (mean, semiamplitude, lower and upper limit) are listed for
each case in the lower panel.   See text for further details.}
\label{t:flowr}
\begin{center}
\begin{tabular}{llllll} \hline
\multicolumn{1}{c}{Q} &
\multicolumn{1}{c}{XY} &
\multicolumn{1}{c}{$(F_{\rm Q})_{\rm XY}$} &
\multicolumn{1}{c}{$\sigma_{(F_{\rm Q})_{\rm XY}}$} &
\multicolumn{1}{c}{$(F_{\rm Q})_{\rm XY}^-$} &
\multicolumn{1}{c}{$(F_{\rm Q})_{\rm XY}^+$} \\
\hline
O              & HH & 0.3871 & 0.1851 & 0.0169 & 0.7573 \\
               & KD & 1.2899 & 0.4445 & 0.4009 & 2.1789 \\
               & HA & 0.3799 & 0.1919 & 0      & 0.7637 \\
Na             & HH & 0.2723 & 0.1315 & 0.0093 & 0.5353 \\
               & KD & 0.3776 & 0.1424 & 0.0928 & 0.6624 \\
               & HA & 0.2477 & 0.1057 & 0.0363 & 0.4591 \\
Mg             & HH & 0.2880 & 0.1010 & 0.0860 & 0.4900 \\
               & KD & 0.5811 & 0.1720 & 0.2371 & 0.9251 \\
               & HA & 0.2824 & 0.1264 & 0.0296 & 0.5352 \\
Si             & HH & 0.1979 & 0.0925 & 0.0129 & 0.3829 \\
               & KD & 0.3853 & 0.1420 & 0.1013 & 0.6693 \\
               & HA & 0.1979 & 0.0925 & 0.0129 & 0.3829 \\
Ca             & HH & 0.4034 & 0.2314 & 0      & 0.8662 \\
               & KD & 1.0722 & 0.5604 & 0      & 2.1930 \\
               & HA & 1.0704 & 0.6017 & 0      & 2.2738 \\
Ti             & HH & 0.5058 & 0.2684 & 0      & 1.0426 \\
               & KD & 1.1454 & 0.6188 & 0      & 2.3830 \\
               & HA & 0.7058 & 0.4057 & 0      & 1.5172 \\
Cr             & HH & 0.5121 & 0.1503 & 0.2115 & 0.8127 \\
               & KD & 0.8363 & 0.2753 & 0.2857 & 1.3869 \\
               & HA & 0.7249 & 0.1887 & 0.3475 & 1.1023 \\
Fe             & HH & 0.6313 & 0.2721 & 0.0871 & 1.1755 \\
               & KD & 0.7726 & 0.4404 & 0      & 1.6534 \\
               & HA & 0.5616 & 0.3245 & 0      & 1.2106 \\
Ni             & HH & 0.3603 & 0.1465 & 0.0673 & 0.6533 \\
               & KD & 0.5804 & 0.1359 & 0.3086 & 0.8522 \\
               & HA & 0.3585 & 0.1305 & 0.0975 & 0.6195 \\
               &    &        &        &        &        \\
$\cap_{\rm Q}$ & HH & 0.2972 & 0.0852 & 0.2115 & 0.3829 \\
               & KD & 0.5317 & 0.1308 & 0.4009 & 0.6625 \\
               & HA & 0.3652 & 0.0177 & 0.3475 & 0.3829 \\
\hline
\end{tabular}                     
\end{center}                      
\end{table*}                       

For an assigned population, the flow parameter, $\kappa$, must necessarily
remain unchanged for different elements.
The intersection of assumed validity ranges, $(F_{\rm Q})_{\rm XY}^\mp=
(F_{\rm Q})_{\rm XY}\mp2\sigma_{(F_{\rm Q})_{\rm XY}}$, is denoted as
$\cap_{\rm Q}$ and related values (mean, semiamplitude, lower and upper limit)
are listed for each case in the bottom panel of Table \ref{t:flowr}.   The
last results disclose that, with respect to LH population environment, HH, KD,
HA population environments were characterized by an outflow to star formation
rate ratio lower than about 30\%, 53\%, 37\%, respectively.

The above considerations hold within the framework of simple MCBR models of
chemical
evolution, which imply (among others) the assumption of instantaneous
mixing.   An opposite extreme situation may be the following: chemical
enrichment took place before sample stars were formed, then abundance
differences are entirely due to cosmic scatter.   If cosmic scatter obeys a
Gaussian distribution where the mean and the variance can be evaluated from
the data, the
theoretical differential abundance distribution reads (Caimmi 2013):
\begin{lefteqnarray}
\label{eq:pscs}
&& (\psi)_{\rm cs}=\log\left\{\frac1{\ln10}\frac1{\sqrt{2\pi}\sigma_{\rm Q}}
\exp\left[-\frac{(\log\phi-\overline{\log\phi_{\rm Q}})^2}{2\sigma_
{\rm Q}^2}\right]\frac1{\phi_{\rm Q}}\right\}~~;
\end{lefteqnarray}
where the index, cs, denotes cosmic scatter, $\overline{\log\phi_{\rm Q}}=
\overline{{\rm [Q/H]}}$ and $\sigma_{\rm Q}=\sigma_{\rm [Q/H]}$.   Related
curves, expressed by Eq.\,(\ref{eq:pscs}), are plotted in
Figs.\,\ref{f:Ohlk4d}-\ref{f:Nilhk4d} for Q = O, Na, Mg, Si, Ca, Ti, Cr, Fe,
Ni, with regard to LH, HH, KD, HA subsamples.   An inspection of 
Figs.\,\ref{f:Ohlk4d}-\ref{f:Nilhk4d}, leaving aside bins populated by a
single star when appropriate, discloses the following.
\begin{description}
\item[(1)\hspace{2.0mm}]
The declining part of the distribution, covering a metallicity range where
most data lie, is slightly different from a straight line.
\item[(2)\hspace{2.0mm}]
The rising part of the distribution, in principle, offers a natural solution
to the FGK-dwarf problem.
\item[(3)\hspace{2.0mm}]
The distribution (full curve) fits to the data to a comparable extent with
respect to simple MCBR models (dashed straight line) with a
slight preference for the last alternative in a few cases, for LH, HH, KD
subsamples.   The same holds for HA subsample, with the exception of O, Si,
Ca, where a marked preference towards chemical evolution, within the framework
of simple MCBR models, can be recognized.
\end{description}
In conclusion, both instantaneous mixing (in the framework of MCBR models) and
cosmic scatter offer viable interpretations to the data, provided HH and KD
populations underwent distinct chemical evolution, as suggested by a recent
investigation (Ishigaki et al. 2013).   If the contrary holds and HA
population is considered, then instantaneous mixing in presence of chemical
evolution is preferred with respect to cosmic scatter in absence of chemical
evolution.

Different kinematical trends have been recently found for LH, HH, KD
populations.   More specifically, LH stars show larger Galactocentric distance
components on both the equatorial plane and the polar axis, with respect to
HH and KD stars.   Accordingly, LH and HH stars may be conceived as
requiring different formation scenarios, with LH stars being accreted.   For
further details refer to the parent paper (Schuster et al. 2012).

Lower [O/Fe] values shown by LH stars with respect to HH stars may
be explained in different ways according if the variation is in [O/H] or in
[Fe/H] or both.   Lower [O/H] values in LH stars could be related to oxygen
depletion in second generation stars within globular clusters, while higher
[Fe/H] values could be related to the contribution from SNIa explosions and
subsequent star formation, regardless of the birth place.   In both cases, HH
population appears to be older
than LH population, which implies similar kinematical trends if the above
mentioned populations formed in situ, contrary to
current data (Ra12).

An interpretation in the framework of the secondary infall scenario could be
the following.   The environment of HH population is related to the inner and
denser region of the proto-Galaxy, which first virialized while the external
shells were still expanding.   The environment of LH population is related to
the outer and less dense region of the proto-Galaxy, which virialized at a
later epoch and probably mixed with SNIa ejecta before forming the first star
generation.

\section{Conclusion}
\label{conc}

A linear [Q/H]-[O/H] relation has been inferred from different populations
sampled in recent attempts (NS10; Ra12), namely LH (low-$\alpha$ halo stars,
$N=24$); HH (high-$\alpha$ halo stars, $N=25$); KD (thick disk stars, $N=16$);
for Q = Na, Mg, Si, Ca, Ti, Cr, Fe, Ni.

The empirical, differential element abundance distribution has been determined
for different populations together with related theoretical counterpart within
the framework of simple MCBR models.   Fractional yields have been inferred
from the data in the framework of simple MCBR models, including an example of
comparison with theoretical counterparts deduced from SNII progenitor
nucleosynthesis for solar and subsolar metallicity, under the assumption of
power-law stellar initial mass function.

Regardless of the chemical evolution model, fractional generalized yields
have been determined.   The ratio of outflow to star formation rate has
been evaluated for a selected population with respect to a reference one.

The theoretical, differential element abundance distribution has been inferred
from the data for different populations, in the opposite limit of
inhomogeneous mixing due to cosmic scatter obeying a Gaussian distribution
whose mean and variance have been evaluated from the related subsample.

The main results may be summarized as follows.
\begin{description}
\item[(1)\hspace{2.0mm}]
With regard to the (${\sf O}$[O/H][Q/H]) plane, stars display along a ``main
sequence'', [Q,O] = $[a_{\rm Q},b_{\rm Q},\Delta b_{\rm Q}]$,
in connection with the straight line, [Q/H] = $a_{\rm Q}$[O/H]+$b_{\rm Q}$.
For unit slopes, $a_{\rm Q}=1$, a main sequence relates to constant [O/Q]
abundance ratio.   In most cases (e.g., Na) stars from OL subsample (two
globular cluster outliers) lie outside the main sequence.
\item[(2)\hspace{2.0mm}]
Regardless of the population, regression line slope estimators fit to the
unit slope within $\mp2\hat\sigma_{\hat a_{\rm Q}}$ for Mg, Si, Ti; within
$\mp3\hat\sigma_{\hat a_{\rm Q}}$ for Cr, Fe, Ni; within $\mp r\hat\sigma_
{\hat a_{\rm Q}}$, $r>3$, for Ca, Na;
where the fit to the unit slope implies related elements are simple primary
i.e. synthesised within SNII progenitors in presence of universal
stellar initial mass function.
\item[(3)\hspace{2.0mm}]
Within the framework of simple MCBR chemical evolution models (Caimmi 
2011a; 2012a), fractional yields are consistent with theoretical results
from SNII progenitor nucleosynthesis (Woosley and Weaver 1995) for Na, Mg,
Ca, Ni, while the contrary holds for the remaining elements in connection with
one subsample at least.   In particular, for Ti,
Cr, Fe, theoretical values appear to be understimated but the
contribution from SNIa progenitors could fill the gap.
\item[(4)\hspace{2.0mm}]
Within the framework of simple MCBR models, a ratio of outflow to star
formation rate has been inferred, of about 30\%, 53\%, 37\%, for HH, KD, HA
population environment, respectively, in comparison with LH population
environment.
\item[(5)\hspace{2.0mm}]
Theoretical, differential element abundance distributions due to cosmic
scatter obeying a Gaussian distribution, fit to the data to a comparable
extent with respect to its counterpart within the framework of simple MCBR
models, for LH, HH, KD population, while the last alternative is preferred for
HA population provided the inner halo and the thick disk underwent common
chemical evolution.
\end{description}

\section*{Acknowledgement}
Thanks are due to the referee, S. Ninkovi\'c, for critical comments which
improved an earlier version of the current paper.

\appendix
\section*{Appendix}

\section{Solar photospheric mass abundances}\label{a:sola}

Solar photospheric mass abundances may be inferred from the following general
relations:
\begin{lefteqnarray}
\label{eq:ZQ}
&& Z_{\rm Q}=\frac{M_{\rm Q}}M=\frac{M_{\rm Q}}{M_{\rm H}}\frac{M_{\rm H}}M=
\frac{N_{\rm Q}\overline{m}_{\rm Q}}{N_{\rm H}\overline{m}_{\rm H}}X~~; \\
\label{eq:Q}
&& {\cal Q}=12+\log\left(\frac{N_{\rm Q}}{N_{\rm H}}\right)~~; \\
\label{eq:mQH}
&& \frac{\overline{m}_{\rm Q}}{\overline{m}_{\rm H}}=
\frac{\sum_kP_{\rm Q_k}m_{\rm Q_k}}{\sum_jP_{\rm H_j}m_{\rm H_j}}=
\frac{\sum_kP_{\rm Q_k}A_{\rm Q_k}}{\sum_jP_{\rm H_j}A_{\rm H_j}}~~;
\end{lefteqnarray}
where $\overline{m}_{\rm Q}$ is the mean atomic mass of the element, Q, in
units of the proton mass, $m_p$; ${\cal Q}$ is an indicator of the fractional
number abundance of the element, Q, with respect to hydrogen;
$P_{\rm Q_k}$ is the fractional abundance of
the isotopic species, ${\rm Q_k}$ (Q = H for hydrogen), $\sum_kP_{\rm Q_k}=1$;
$A_{\rm Q_k}$ is the mass number of the isotopic species, ${\rm Q_k}$.   The
result is:
\begin{lefteqnarray}
\label{eq:ZQXs}
&& \frac{(Z_{\rm Q})_\odot}{X_\odot}=\exp_{10}({\cal Q}-12)
\frac{\sum_kP_{\rm Q_k}A_{\rm Q_k}}{\sum_jP_{\rm H_j}A_{\rm H_j}}~~;
\end{lefteqnarray}
which can be inserted into Eq.\,(\ref{eq:CQ}).   The results for the solar
photospheric mass abundances, $Z_{\rm Q}$, Q = O, Na, Mg, Si, Ca, Ti, Cr, Fe,
Ni, are listed in Table \ref{t:solab}.
\begin{table}
\caption{Solar photospheric mass abundances, $Z_{\rm Q}$, inferred from
number abundance indicators, ${\cal Q}$ (normalized to ${\cal H}=12$),
and isotopic abundance fractions of the solar photosphere
determined in a recent investigation (Asplund et al. 2009,
Tables 1, 3, 4, therein).   Hydrogen abundance, $Z_{\rm H}=X$,
is necessary for evaluating $Z_{\rm Q}$, Q $\ne$ H, according to
Eq.\,(\ref{eq:ZQXs}). The atomic number is denoted as ${\cal Z}$.   For
further details refer to the text.}
\label{t:solab}
\begin{center}
\begin{tabular}{rlrl}
\multicolumn{1}{c}{${\cal Z}$} &
\multicolumn{1}{l}{Q} &
\multicolumn{1}{c}{${\cal Q}$} &
\multicolumn{1}{c}{$Z_{\rm Q}$} \\
\hline
 1 & H  & 12.00 & 7.381E$-1$ \\
 8 & O  &  8.69 & 5.786E$-3$ \\
11 & Na &  6.24 & 2.950E$-5$ \\
12 & Mg &  7.60 & 7.146E$-4$ \\
14 & Si &  7.51 & 6.713E$-4$ \\
20 & Ca &  6.34 & 6.478E$-5$ \\
22 & Ti &  4.95 & 3.152E$-6$ \\
24 & Cr &  5.64 & 1.677E$-5$ \\
26 & Fe &  7.50 & 1.305E$-3$ \\
28 & Ni &  6.22 & 7.198E$-5$ \\
\hline                            
\end{tabular}                     
\end{center}                      
\end{table}                       
Related values for helium and metals are $Y=Z_{\rm He}=0.2485$ and
$Z=\sum_{{\rm Q}\ne{\rm H, He}}Z_{\rm Q}=0.0134$, respectively (Asplund et al.
2009).

\section{Fractional yields in simple MCBR models}\label{a:fry}

With regard to simple MCBR chemical evolution models (Caimmi 2011a; 2012),
the combination of Eqs.\,(\ref{eq:phiQ}) and (\ref{eq:phiO}) yields:
\begin{equation}
\label{eq:phQO}
\frac{\phi_{\rm Q}-(\phi_{\rm Q})_i}{\phi_{\rm O}-(\phi_{\rm O})_i}=\frac
{(Z_{\rm O})_\odot}{(Z_{\rm Q})_\odot}\frac{\hat{p}_{\rm Q}}{\hat{p}_{\rm O}}
~~;
\end{equation}
where, on the other hand, an assumption of the model is $Z=c_{\rm Q}
Z_{\rm Q}=c_{\rm O}Z_{\rm O}$, $Z$ global fractional metal mass abundance,
$c_{\rm Q}$ and $c_{\rm O}$ constant,
(Caimmi 2011a), which via Eq.\,(\ref{eq:fQfH}) is equivalent to:
\begin{equation}
\label{eq:rfQO}
\frac{\phi_{\rm Q}}{\phi_{\rm O}}=\frac{(\phi_{\rm Q})_i}{(\phi_{\rm O})_i}=
\frac{c_{\rm O}}{c_{\rm Q}}\frac{(Z_{\rm O})_\odot}{(Z_{\rm Q})_\odot}~~;
\end{equation}
provided Q and O are simple primary elements.

Starting from $Z-Z_i=c_{\rm Q}[Z_{\rm Q}-(Z_{\rm Q})_i]=c_{\rm O}[Z_{\rm O}-
(Z_{\rm O})_i]$ and following the same way yields:
\begin{equation}
\label{eq:rdQO}
\frac{\phi_{\rm Q}-(\phi_{\rm Q})_i}{\phi_{\rm O}-(\phi_{\rm O})_i}=
\frac{c_{\rm O}}{c_{\rm Q}}\frac{(Z_{\rm O})_\odot}{(Z_{\rm Q})_\odot}~~;
\end{equation}
and the combination of Eqs.\,(\ref{eq:rfQO}) and (\ref{eq:rdQO}) produces:
\begin{equation}
\label{eq:fdQO}
\frac{\phi_{\rm Q}-(\phi_{\rm Q})_i}{\phi_{\rm O}-(\phi_{\rm O})_i}=
\frac{\phi_{\rm Q}}{\phi_{\rm O}}~~;
\end{equation}
which implies $\phi_{\rm Q}/\phi_{\rm O}=(\phi_{\rm Q})_i/(\phi_{\rm O})_i$,
as expected.

Finally, the substitution of Eq.\,(\ref{eq:fdQO}) into (\ref{eq:phQO}) yields
Eq.\,(\ref{eq:fy}).

\section{Fractional yield, intercept and fractional slope uncertainties}
\label{a:err}

Fractional yield, intercept and fractional slope uncertainties, mentioned in
the text, are evaluated using standard formulae of error propagation.   Though
only quadratic errors have been used in the current attempt, for sake of
completeness also absolute errors shall be included in the following.

Let $m_1$, $m_2$, ..., $m_v$, be independent random variables obeying  
Gaussian
distributions and let $m$ be a random variable which depends on $m_1$, $m_2$,
$...$, $m_v$, as $m=f(m_1, m_2, ..., m_v,)$, where $f$ is a specified
continuous and derivable function.   According to a theorem of statistics,
related quadratic and absolute errors read:
\begin{lefteqnarray}
\label{eq:sgm}
&& \sigma_m=\left\{\sum_{r=1}^v\left[\left(\frac{\partial f}{\partial m_r}
\right)_{{\sf P}^\ast}\sigma_{m_r}\right]^2\right\}^{1/2}~~; \\
\label{eq:aem}
&& \Delta m=\sum_{r=1}^v\left\vert\left(\frac{\partial f}
{\partial m_r}\right)_{{\sf P}^\ast}\right\vert\Delta m_r~~;
\end{lefteqnarray}
where ${\sf P}^\ast\equiv(m_1^\ast, m_2^\ast, ..., m_v^\ast)$, $m_r^\ast$ is
the expected value
(approximated by the mean) of the distribution depending on $m_r$,
$\sigma_{m_r}$ the related rms error (approximated by the rms deviation),
$\Delta m_r$ the maximum error (in absolute value) on the determination of
$m_r$.

The particularization of Eqs.\,(\ref{eq:sgm}), (\ref{eq:aem}), to the
fractional yield, $\hat{p}_{\rm Q}/\hat{p}_{\rm O}$, expressed by
Eq.\,(\ref{eq:fg}), after some algebra yields:
\begin{lefteqnarray}
\label{eq:sgQOb}
&& \sigma_{\hat{p}_{\rm Q}/\hat{p}_{\rm O}}=\frac{\hat{p}_{\rm Q}}
{\hat{p}_{\rm O}}\frac{\sigma_{b_{\rm Q}}}{\ln10}~~; \\
\label{eq:aeQOb}
&& \Delta\frac{\hat{p}_{\rm Q}}{\hat{p}_{\rm O}}=\frac{\hat{p}_{\rm Q}}
{\hat{p}_{\rm O}}\frac{\Delta b_{\rm Q}}{\ln10}~~;
\end{lefteqnarray}
where $\sigma_{b_{\rm Q}}$, $\Delta b_{\rm Q}$, can be inferred from Table
\ref{t:rette}.

The particularization of Eqs.\,(\ref{eq:sgm}), (\ref{eq:aem}), to the
fractional yield, $\hat{p}_{\rm Q}/\hat{p}_{\rm O}$, expressed by
Eq.\,(\ref{eq:alOQ}), after some algebra yields:
\begin{lefteqnarray}
\label{eq:sgQOa}
&& \sigma_{\hat{p}_{\rm Q}/\hat{p}_{\rm O}}=\frac{\hat{p}_{\rm Q}}
{\hat{p}_{\rm O}}\left[\left(\frac{\sigma_{\alpha_{\rm O}}}{\alpha_{\rm O}}
\right)^2+\left(\frac{\sigma_{\alpha_{\rm Q}}}{\alpha_{\rm Q}}\right)^2\right]
^{1/2}~~; \\
\label{eq:aeQOa}
&& \Delta\frac{\hat{p}_{\rm Q}}{\hat{p}_{\rm O}}=\frac{\hat{p}_{\rm Q}}
{\hat{p}_{\rm O}}\left[\left\vert\frac{\Delta\alpha_{\rm O}}{\alpha_{\rm O}}
\right\vert+\left\vert\frac{\Delta\alpha_{\rm Q}}{\alpha_{\rm Q}}\right\vert
\right]~~;
\end{lefteqnarray}
where $\sigma_{\alpha_{\rm Q}}$, $\Delta \alpha_{\rm Q}$, can be inferred from
Table \ref{t:rettd}.

The particularization of Eqs.\,(\ref{eq:sgm}), (\ref{eq:aem}), to the
intercept, $b_{\rm Q}$, expressed by Eq.\,(\ref{eq:bQ}), after some algebra
yields:
\begin{lefteqnarray}
\label{eq:sgbQ}
&& \sigma_{b_{\rm Q}}=\frac1{\ln10}\left[\left(\frac{\sigma_{\alpha_{\rm O}}}
{\alpha_{\rm O}}\right)^2+\left(\frac{\sigma_{\alpha_{\rm Q}}}{\alpha_{\rm Q}}
\right)^2\right]^{1/2}~~; \\
\label{eq:aebQ}
&& \Delta b_{\rm Q}=\frac1{\ln10}\left[\left\vert\frac{\Delta\alpha_{\rm O}}
{\alpha_{\rm O}}\right\vert+\left\vert\frac{\Delta\alpha_{\rm Q}}
{\alpha_{\rm Q}}\right\vert\right]~~;
\end{lefteqnarray}
where $\sigma_{\alpha_{\rm Q}}$, $\Delta \alpha_{\rm Q}$, can be inferred from
Table \ref{t:rettd}.

The particularization of Eqs.\,(\ref{eq:sgm}), (\ref{eq:aem}), to the
fractional slope, $(\alpha_{\rm Q})_{\rm P1}/(\alpha_{\rm Q})_{\rm P2}$, after
some algebra yields:
\begin{lefteqnarray}
\label{eq:saQOb}
&& \sigma_{(\alpha_{\rm Q})_{\rm P1}/(\alpha_{\rm Q})_{\rm P2}}=\frac
{(\alpha_{\rm Q})_{\rm P1}}{(\alpha_{\rm Q})_{\rm P2}}\left[\left(\frac
{(\sigma_{\alpha_{\rm Q}})_{\rm P1}}{(\alpha_{\rm Q})_{\rm P1}}\right)^2+
\left(\frac{(\sigma_{\alpha_{\rm Q}})_{\rm P2}}{(\alpha_{\rm Q})_{\rm P2}}
\right)^2\right]^{1/2}~~; \\
\label{eq:aaQOb}
&& \Delta\frac{(\alpha_{\rm Q})_{\rm P1}}{(\alpha_{\rm Q})_{\rm P2}}=\frac
{(\alpha_{\rm Q})_{\rm P1}}{(\alpha_{\rm Q})_{\rm P2}}\left[\left\vert\frac
{\Delta(\alpha_{\rm Q})_{\rm P1}}{(\alpha_{\rm Q})_{\rm P1}}\right\vert+\left
\vert\frac{\Delta(\alpha_{\rm Q})_{\rm P2}}{(\alpha_{\rm Q})_{\rm P2}}\right
\vert\right]~~;
\end{lefteqnarray}
where $\sigma_{(\alpha_{\rm Q})_{{\rm P}i}}$, $\Delta(\alpha_{\rm Q})_
{{\rm P}i}$, $i=1,2$, can be inferred from Table \ref{t:rettd}.

\section{Fractional yields from star nucleosynthesis}
\label{a:try}

Let $\Delta(m_i)_{\rm Q}$ be the mass in the element, Q, synthesised within a
star of initial mass, $m_i$, $1\le i\le n$, and returned to the interstellar
medium after star death, for which the result is known.   The restriction of
a linear trend between contiguous values, $m_j$, $m_{j+1}$, $1\le j\le n-1$,
reads:
\begin{lefteqnarray}
\label{eq:linj}
&& \frac{\Delta m_{\rm Q}-\Delta(m_j)_{\rm Q}}
{\Delta(m_{j+1})_{\rm Q}-\Delta(m_j)_{\rm Q}}=\frac{m-m_j}{m_{j+1}-m_j}~~;
\end{lefteqnarray}
where $m_j\le m\le m_{j+1}$ without loss of generality.

The straight line, defined by Eq.\,(\ref{eq:linj}), takes the expression:
\begin{lefteqnarray}
\label{eq:DmQ}
&& \Delta m_{\rm Q}=(A_j)_{\rm Q}m+(B_j)_{\rm Q}m_\odot~~; \\
\label{eq:AjQ}
&& (A_j)_{\rm Q}=\frac{\Delta(m_{j+1})_{\rm Q}-\Delta(m_j)_{\rm Q}}
{m_{j+1}-m_j}~~; \\
\label{eq:AjQ}
&& (B_j)_{\rm Q}=\frac{\Delta(m_j)_{\rm Q}}{m_\odot}-(A_j)_{\rm Q}\frac{m_j}
{m_\odot}~~;
\end{lefteqnarray}
where masses are expressed in solar units.

The further restriction of a power-law stellar initial mass function:
\begin{lefteqnarray}
\label{eq:IMF}
&& \phi\left(\frac m{m_\odot}\right)=C\left(\frac m{m_\odot}\right)^{-p}~~;
\end{lefteqnarray}
where $C$ is a normalization constant and $-p$ the power-law exponent, allows
a simple expression for the mass in the element, Q, synthesised within stars
of initial mass, $m_j\le m\le m_{j+1}$, $1\le j\le n-1$, and returned to the
interstellar medium after star death.

The result is:
\begin{lefteqnarray}
\label{eq:DjmQ}
&& \Delta_jm_{\rm Q}=\int_{m_j/m_\odot}^{m_{j+1}/m_\odot}\Delta m_{\rm Q}~
\phi\left(\frac m{m_\odot}\right)\diff\left(\frac m{m_\odot}\right)~~;
\end{lefteqnarray}
and the substitution of Eqs.\,(\ref{eq:DmQ})-(\ref{eq:IMF}) into
(\ref{eq:DjmQ}) after some algebra yields:
\begin{lefteqnarray}
\label{eq:DjeQ}
&& \Delta_jm_{\rm Q}=Cm_\odot\left\{\frac{(A_j)_{\rm Q}}{2-p}\left[\left(\frac
{m_{j+1}}{m_\odot}\right)^{2-p}-\left(\frac{m_j}{m_\odot}\right)^{2-p}\right]
\right.\nonumber \\
&& \phantom{\Delta_jm_{\rm Q}=Cm\,}+\left.
\frac{(B_j)_{\rm Q}}{1-p}\left[\left(\frac{m_{j+1}}{m_\odot}\right)^{1-p}-
\left(\frac{m_j}{m_\odot}\right)^{1-p}\right]\right\}~~;
\end{lefteqnarray}
where the power-law exponent may safely be assumed as lying within the range,
$-3\le-p\le-2$.

The mass in the element, Q, synthesised witin the whole stellar generation
(sg) and returned to the interstellar medium after star death, is:
\begin{lefteqnarray}
\label{eq:DsgmQ}
&& \Delta_{\rm sg}m_{\rm Q}=\sum_{j=1}^{n-1}\Delta_jm_{\rm Q}~~;
\end{lefteqnarray}
where, after substitution of Eq.\,(\ref{eq:DjeQ}) into (\ref{eq:DsgmQ}),
$m_1$ and $m_n$ are the lower and upper mass limit, respectively, of stars
which produce and, when dying, return the element, Q, to the interstellar
medium.

In the framework of simple MCBR models, the yield of the element, Q, can be
expressed as (e.g., Caimmi 2007):
\begin{lefteqnarray}
\label{eq:pQ}
&& \hat{p}_{\rm Q}=\frac{I_{\rm Q}(12)}{\alpha I(1)}~~;
\end{lefteqnarray}
where $I_{\rm Q}(12)=\Delta_{\rm sg}m_{\rm Q}$ and $\alpha$, $I(1)$, are
independent of Q.   Accordingly, the fractional yield related to selected
elements, Q$_1$, Q$_2$, reads:
\begin{lefteqnarray}
\label{eq:ty}
&& \frac{\hat{p}_{\rm Q_1}}{\hat{p}_{\rm Q_2}}=\frac{\Delta_{\rm sg}m_
{\rm Q_1}}{\Delta_{\rm sg}m_{\rm Q_2}}~~;
\end{lefteqnarray}
where the substitution of Eqs.\,(\ref{eq:DjeQ}) and (\ref{eq:DsgmQ}) into
(\ref{eq:ty}) implies the disappearence of the product, $Cm_\odot$.

\end{document}